 \documentclass[prd,twocolumn,showpacs,preprintnumbers,superscriptaddress]{revtex4-1}

\usepackage{upgreek}
\usepackage{siunitx}
\usepackage{enumitem}
\makeatletter
\let\orig@Itemize =\itemize
\let\orig@Enumerate =\enumerate
\let\orig@Description =\description
\renewenvironment{itemize}{\orig@Itemize[noitemsep,nolistsep]}{\endlist}
\renewenvironment{enumerate}{\orig@Enumerate[noitemsep,nolistsep]}{\endlist}
\renewenvironment{description}{\orig@Description[noitemsep,nolistsep]}{\endlist}
\makeatother

\usepackage{xcolor} 
\usepackage{graphicx}
\graphicspath{{/.}{Fig/}} 

\usepackage{mathtools} 
\usepackage{amssymb}
\usepackage{amsfonts}
\usepackage{mathrsfs}

\usepackage{multirow}
\usepackage{dcolumn}
\usepackage{url}
\usepackage{hyperref}

\usepackage[pagewise,mathlines]{lineno}

\newcommand*{\ud}{\ensuremath{\,\mathrm{d}}}
\newcommand*{\tripoli}{TRIPOLI-4\textregistered{}}
\newcommand*{\neb}{\ensuremath{\overline{\nu}_{e}}}

\newcommand*{\Delrel}{\ensuremath{\Delta t_{e^+n}}}
\newcommand*{\Vc}{\ensuremath{\mathscr{V}_{\mathrm{c}}}}
\newcommand*{\Vd}{\ensuremath{\mathscr{V}_{\mathrm{d}}}}
\newcommand*{\rc}{\ensuremath{\overrightarrow{r_{\mathrm{c}}}}}
\newcommand*{\rd}{\ensuremath{\overrightarrow{r_{\mathrm{d}}}}}
\newcommand*{\leff}{\ensuremath{l_{\mathrm{eff}}}}
\newcommand*{\anuen}{\ensuremath{\mathrm{E}_{\overline{\nu}_e}}}

\newcommand*{\Ep}{\ensuremath{\mathrm{E}_{\text{prompt}}}}
\newcommand*{\Ed}{\ensuremath{\mathrm{E}_{\text{delay}}}}
\newcommand{\nuc}[2]{\ensuremath{{}^{#2}\text{#1}}}
\newcommand*{\Urfive}{\ensuremath{^{235}\text{U}}}
\newcommand*{\Ureight}{\ensuremath{^{238}\text{U}}}
\newcommand*{\Punine}{\ensuremath{^{239}\text{Pu}}}
\newcommand*{\Puone}{\ensuremath{^{241}\text{Pu}}}
\newcommand*{\Li}{$^{9}\text{Li}$}
\newcommand*{\He}{$^{8}\text{He}$}

\newcommand*{\Co}{$^{60}\text{Co}$}
\newcommand*{\Cs}{$^{137}\text{Cs}$}
\newcommand*{\Na}{$^{22}\text{Na}$}
\newcommand*{\qtot}{\ensuremath{Q_\text{tot}}}
\newcommand*{\qtail}{\ensuremath{Q_\text{tail}}}
\newcommand*{\psd}{\qtail{}/\qtot{}}
\newcommand{\Rrod}[1]{\ensuremath{R^{\text{rod#1}}}}
\newcommand{\Trod}[1]{\ensuremath{T^{\text{rod#1}}}}
\newcommand*{\taunu}{\ensuremath{\tau_{\bar{\nu}}}}
\newcommand*{\taupr}{\ensuremath{\tau_{\text{prompt}}}}
\newcommand*{\taudel}{\ensuremath{\tau_{\text{delay}}}}
\newcommand{\taug}[1]{\ensuremath{\tau_{\text{#1}}}} 
\newcommand{\Num}[1]{\ensuremath{\N_{\text{#1}}}}
\newcommand{\err}[1]{\ensuremath{\delta_{\text{#1}}}} 
\newcommand{\Rnu}[1]{\ensuremath{R_\nu^{\text{#1}}}}
\newcommand*{\Dist}{\SI{7.21(11)}{m}}
\newcommand*{\Vol}{\SI{846.8(70)}{L}}
\newcommand*{\MeanPower}{\SI{66.5}{MW}}
\newcommand*{\AbsErrEffPC}{\SI{2.2}{\%}}
\newcommand*{\RelErrEffPC}{\SI{7.2}{\%}}
\newcommand*{\EffPC}{\SI{30.3}{\%}}

\newcommand*{\EffwithErrPC}{\SI{30.3(22)}{\%}}
\newcommand*{\RatePred}{\SI{277(23)}{\neb/day}}
\newcommand*{\RateNuMCnoEffConfigA}{\SI{910.8}{\neb/day}}
\newcommand*{\RateNuMCnoEffConfigB}{\SI{913.8}{\neb/day}}
\newcommand*{\RateAccOFFnum}{\num{69.1 \pm 0.1}} 
\newcommand*{\RateCandOFFnum}{\num{1223.5 \pm 3.4}} 
\newcommand*{\RateCorrOFFnum}{\num[parse-numbers=false]{1145.4 \pm 3.4 \pm 2.5}} 

\newcommand*{\RateAccOFF}{\SI[parse-numbers=false]{\RateAccOFFnum{}}{events/day}} 
\newcommand*{\RateCorrOFF}{\SI[parse-numbers=false]{\RateCorrOFFnum{}}{events/day}} 
\newcommand*{\RateAccONnum}{\num{3476.3 \pm 0.7}} 
\newcommand*{\RateCandONnum}{\num{4903 \pm 7}} 
\newcommand*{\RateCorrONnum}{\num[parse-numbers=false]{1426 \pm 7 \pm 18}} 

\newcommand*{\RateAccON}{\SI[parse-numbers=false]{\RateAccONnum{}}{events/day}} 
\newcommand*{\RateObs}{\SI[parse-numbers=false]{281 \pm 7 (stat) \pm 18 (syst)}{\neb/day}} 
\newcommand*{\AccToSignalRatio}{\num{11.9}} 
\newcommand*{\CorrToSignalRatio}{\num{3.9}} 
\newcommand*{\RateObsToPredRatio}{\num{1.014(108)}}

\newcommand*{\CEAirfu}{Commissariat \`{a} l'\'energie atomique et aux \'energies alternatives, Centre de Saclay,
DSM/IRFU, 91191 Gif-sur-Yvette, France}
\newcommand*{\CEAdam}{Commissariat \`{a} l'\'energie atomique et aux \'energies alternatives, DAM, DIF, 91297, Arpajon, France}
\newcommand*{\CEAden}{Commissariat \`{a} l'\'energie atomique et aux \'energies alternatives, Centre de Saclay,
DEN, 91191 Gif-sur-Yvette, France}
\newcommand*{\MaxPlanck}{Max-Planck-Institut f\"{u}r Kernphysik, 69029 Heidelberg, Germany}
\newcommand*{\SUBATECH}{SUBATECH, CNRS/IN2P3, Universit\'e de Nantes, Ecole des Mines de Nantes, F-44307 Nantes, France}

\begin{document}

\author{G.~Boireau}
\author{L.~Bouvet}
\author{A.P.~Collin}
\author{G.~Coulloux}
\author{M.~Cribier}
\author{H.~Deschamp}
\author{V.~Durand}
\author{M.~Fechner}
\author{V.~Fischer}
\author{J.~Gaffiot}
\author{N.~G\'erard Castaing}
\author{R.~Granelli}
\author{Y.~Kato}
\author{T.~Lasserre}
\author{L.~Latron}
\author{P.~Legou}
\author{A.~Letourneau}
\author{D.~Lhuillier}
\author{G.~Mention}
\author{Th.~A.~Mueller}
\author{T-A.~Nghiem}
\author{N.~Pedrol}
\author{J.~Pelzer}
\author{M.~Pequignot}
\author{Y.~Piret}
\author{G.~Prono}
\author{L.~Scola}
\author{P.~Starzinski}
\author{M.~Vivier}
\affiliation{\CEAirfu}

\author{E.~Dumonteil}
\author{D.~Mancusi}
\affiliation{\CEAden}

\author{C.~Varignon}
\affiliation{\CEAdam}

\author{C.~Buck}
\author{M.~Lindner}
\affiliation{\MaxPlanck}

\author{J.~Bazoma}
\author{S.~Bouvier}
\author{V.M.~Bui}
\author{V.~Communeau}
\author{A.~Cucoanes}
\author{M.~Fallot}
\author{M.~Gautier}
\author{L.~Giot}
\author{G.~Guilloux}
\author{M.~Lenoir}
\author{J.~Martino}
\author{G.~Mercier}
\author{T.~Milleto}
\author{N.~Peuvrel}
\author{A.~Porta}
\author{N.~Le~Qu\'er\'e}
\author{C.~Renard}
\author{L.M.~Rigalleau}
\author{D.~Roy}
\author{T.~Vilajosana}
\author{F.~Yermia}
\affiliation{\SUBATECH}

\collaboration{The Nucifer Collaboration}
\noaffiliation

\title{Online Monitoring of the Osiris Reactor with the Nucifer Neutrino Detector}

\begin{abstract}
Originally designed as a new nuclear reactor monitoring device, the Nucifer detector has successfully detected
its first neutrinos. We provide the second shortest baseline measurement of the reactor neutrino flux.
The detection of electron antineutrinos emitted in the decay chains of the fission products,
combined with reactor core simulations, provides a new tool to assess both the thermal power
and the fissile content of the whole nuclear core and could be used by the International
Agency for Atomic Energy (IAEA) to enhance the Safeguards of civil nuclear reactors.
Deployed at only \SI{7.2}{m} away from the compact Osiris research reactor core (\SI{70}{MW}) operating
at the Saclay research centre of the French Alternative Energies and Atomic Energy Commission (CEA),
the experiment also exhibits a well-suited configuration to search for a new short baseline oscillation.
We report the first results of the Nucifer experiment, describing the performances of the $\sim \SI{0.85}{m^3}$
detector remotely operating at a shallow depth equivalent to $\sim \SI{12}{m}$ of water
and under intense background radiation conditions. 
Based on 145 (106) days of data with reactor ON (OFF), leading to the detection of
an estimated \SI{40760}{\neb}, the mean number of detected antineutrinos is \RateObs{}, in agreement with
the prediction \RatePred{}. Due to the large background no conclusive results on the existence of light sterile
neutrinos could be derived, however. As a first societal application we quantify how antineutrinos could be used
for the Plutonium Management and Disposition Agreement.

\end{abstract}

\maketitle





\section{Introduction}
\label{sec:intro}


In a context of increasing needs for carbon emission-free energy, civilian nuclear power generation
has the potential to play an important role in global energy production, and the list of countries aiming
to acquire technological know-how in the field of civilian nuclear energy could increase.
As a consequence, the International Atomic Energy Agency (IAEA) has been evaluating the potential of new technologies
to guarantee that nations use nuclear energy only for peaceful purposes.

Neutrino detectors have the unique ability to non-intrusively monitor a nuclear reactor's operational status,
thermal power and fissile content in real-time, from outside the reactor containment. More specifically the scenarios
of interest are to confirm the absence of unrecorded production of fissile materials in declared reactors
as well as to estimate the total burn-up\footnote{Burn-up or fuel utilization is the amount of energy extracted
from a given quantity of nuclear fuel.} of a reactor core.
Nucifer is a detector first built for long term reliable safeguards measurements in the vicinity
of operating nuclear reactor cores. The experiment aims at demonstrating the concept of ``neutrinometry''
at the pre-industrialized stage. Therefore a well-established technology and commercial components
were chosen for the detection system.


In addition, after the re-evaluation of antineutrino spectra~\cite{Mueller:2011,Huber:2011wv,Haag:2013raa},
the reanalysis of all short baseline reactor experiments~\cite{Mention:2011rk} lead to what is known as
the Reactor Antineutrino Anomaly (RAA) with $\Rnu{obs}/\Rnu{pred} = 0.936 \pm 0.024$~\cite{Lasserre:2014}.
And it turns out that the Nucifer experimental set-up is adequate to probe the Reactor Antineutrino
Anomaly~\cite{Mention:2011rk}, searching for possible oscillations into sterile neutrino species at very short baselines.
Indeed, Nucifer is a compact detector ($\sim$ \SI{1.2}{m} in diameter, \SI{0.7}{m} in height), deployed at only \Dist{}
(centre to centre) from a compact nuclear core (\SI{57 x 57 x 60}{cm}). Therefore the three conditions to search
for short baseline oscillations are met. Nucifer is even the second shortest baseline reactor neutrino experiment,
the only closest experiment being a \SI{3.2}{L} Gd-loaded liquid scintillator detector deployed at only \SI{6.5}{m}
from the Savannah River Plant (USA) in 1965~\cite{Nezrick:1966pn,Silverman:1981md}.
The third shortest baseline is the ILL experiment~\cite{Kwon:1981ua} with \SI{377}{L} of liquid scintillator
at \SI{8.76}{m} from the core, which lead to a ratio of experimental to expected integral positron yield of
\SI[parse-numbers=false]{0.832 \pm \SI{3.5}{\%} (stat) \pm \SI{8.87}{\%} (syst)}{} as stated in the 1995
reanalysis~\cite{Hoummada1995}.


Among the possible societal applications the Plutonium Management and Disposition Agreement (PMDA)~\cite{PMDA:2010}
could be monitored through neutrino rate monitoring. Indeed, in this procedure, weapon-grade plutonium
could be processed into Mixed OXide uranium-plutonium (MOX) fuel, irradiated in civil nuclear power reactors,
and therefore transformed into material unusable in the fabrication of nuclear weapons.

The paper first describes the experimental lay-out in part~\ref{sec:ExpLayout}, from the Osiris reactor to the data
acquisition system through the Nucifer detector. We then calculate the expected signal and its associated errors
in part~\ref{sec:Signal}, accounting for the major part of the final uncertainty due to the relatively low detector efficiency.
Then the calibration system is depicted in part~\ref{sec:calib}, followed by the data analysis: we detail the data
sample in section~\ref{sec:DataSample} and the neutrino candidate selection in section~\ref{sec:NuCuts}, then the
accidental background in section~\ref{sec:acc}, the cosmic induced correlated background in section~\ref{sec:cosmic_bkg},
the reactor induced correlated backgrounds in section~\ref{sec:reactor_bkg} and finally the detection efficiency and
the associated uncertainties in section~\ref{sec:DetEffAndSyst}.
We thus present the Nucifer results at Osiris in part~\ref{sec:results}, summarized by the ratio of observed to expected
neutrino detection rate.
We finally discuss a potential application for the PMDA agreement, using Nucifer's collected data, in part~\ref{sec:Pu_mass}.

\section{Experimental layout} 
\label{sec:ExpLayout}
\subsection{Deployment site: the Osiris reactor}
\label{sec:Reactor}

\begin{figure}[h!]
\centering \includegraphics[width=1 \linewidth]{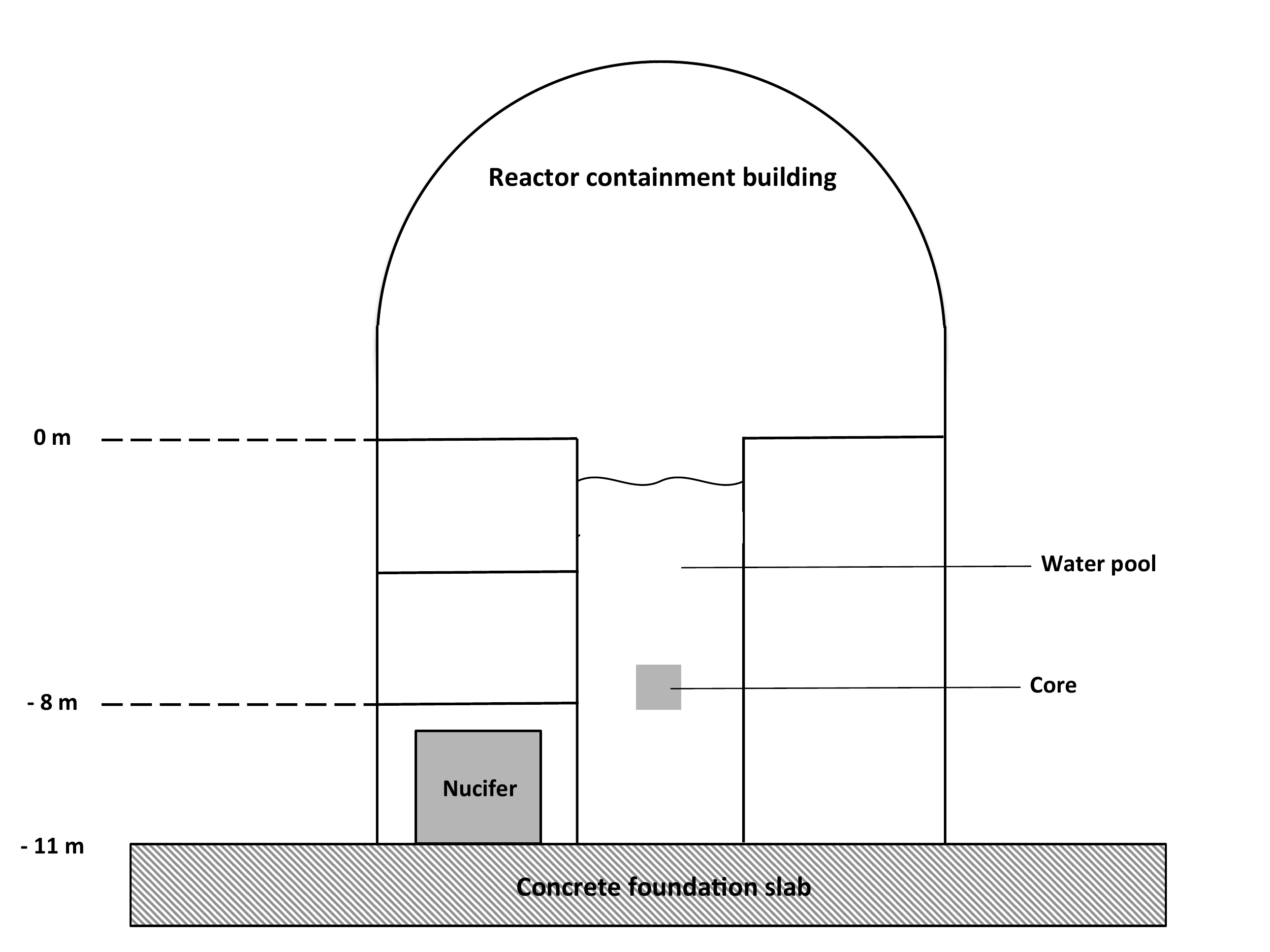}
\caption{The Nucifer experimental layout. The detector centre is located \Dist{} away from
the reactor core centre. East is on the right, south points to the reader.}
\label{fig:exp_layout}
\end{figure}

Nucifer is installed on the concrete foundation slab of the Osiris reactor building, \SI{11}{meters} beneath
the water pool level, in a dedicated room next to the reactor core (see Fig.~\ref{fig:exp_layout}). Such a
location allows to safely support the weight of the detector together with its heavy passive shielding
($\gtrsim \SI{62}{tonnes}$). It also offers a modest overburden, reducing the muon flux by a factor~of~2.7 with
respect to sea level. The overburden is equivalent to $\sim \SI{12}{meters}$ of water.

The reactor core, approximately at the Nucifer room ceiling level, is located \Dist{} away from the detector,
from the centre of the core to the centre of the detector.
In this configuration radiations from the reactor core are attenuated by about \SI{2}{m} of concrete and \SI{3.5}{m} of water.
Nevertheless, the level of gamma radiations in the Nucifer room is still quite high during the reactor operations,
enhancing the challenge of properly extracting the neutrino flux.

Osiris is a light water experimental reactor of open-core pool type located at the Saclay research centre of
the French Alternative Energies and Atomic Energy Commission (CEA). It operates at a nominal thermal power of
\SI{70}{MW} and has been designed for technological irradiation purposes and radioisotope production~\cite{Osiris1}.
The fissile elements produce a high neutron flux in the core, at the level of a few \SI{e14}{neutrons.cm^{-2}.s^{-1}},
both in the thermal and fast energy range. The core size is \SI{57 x 57 x 60}{cm}, excluding its vessel, corresponding
to 56~cells loaded with 38~standard fuel elements ($\text{U}_3\text{Si}_2\text{Al}$ plates enriched
at \SI{19.75}{\%} in \Urfive{}), 6~control elements (made of Hafnium absorber in the upper part and of fuel
in the lower part), 7~Beryllium elements used as neutron reflectors, and 5~cells equipped with water boxes dedicated
to experiments. The cells are arranged in a square lattice with a lattice parameter of \SI{8.74}{cm}.
The absence of pressurization vessel allows for a direct access to the core at any time.

\begin{figure}[h!]
\centering \includegraphics[width=1.\linewidth]{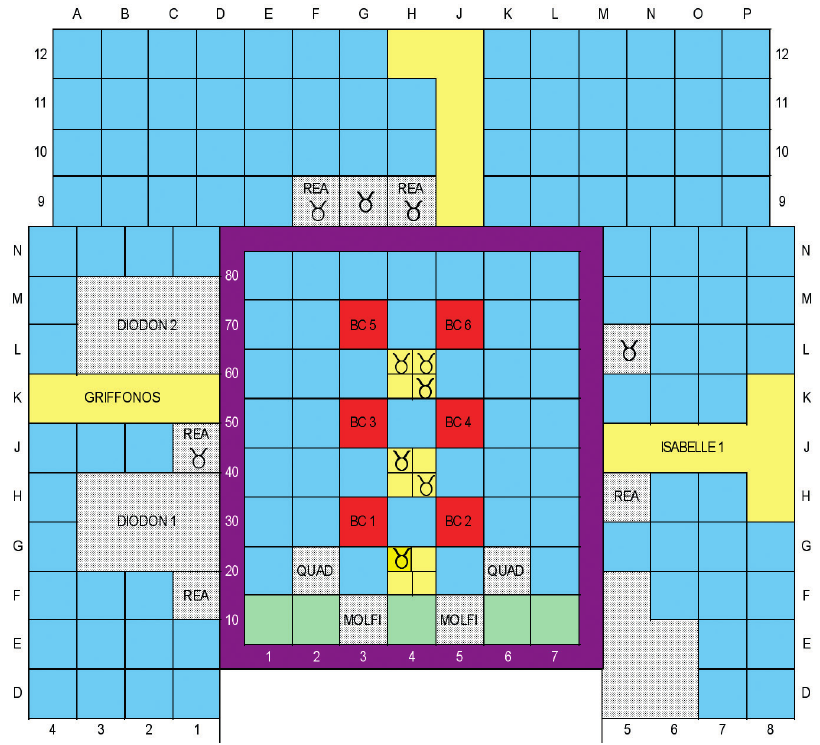}
\caption{Top scheme of the Osiris core.
``BC'' squares (red) stand for core control elements, bottom core line squares (labelled 10, green)
for Beryllium reflector elements.
The square surrounding the core (purple) represents the steel vessel.
Other areas (yellow, grey) stand for locations used for experimental irradiation devices,
or industrial irradiation locations~\cite{Osiris2}.
North is on top, Nucifer is located on the left. }
\label{fig:Osiris}
\end{figure}

Osiris typically operates \SI{180}{days} per year with \SI{3}{week} reactor cycles, core being
refuelled by about~$1/7^{th}$. The \SI{19.75}{\%} enrichment of the nuclear fuel in \Urfive{} and the short cycle
duration suppress the evolution of the isotopic composition of the nuclear fuel.
To start the reactor, control rods 1, 2, 5 and 6 are first raised, and the core reaches criticality during
the raising of the next rod which can be the 3 or the 4 (see fig~\ref{fig:Osiris}). This rod is then raised
to the top within the next few days, and the last rod is progressively raised during the remaining part of the cycle.
This sequence leads to a displacement of the fission barycentre, in the east/west direction and vertically,
different for the two rod configurations (3 then 4 or 4 then 3).

\subsection{Nucifer detector}
\label{sec:Detector}
\begin{figure}[h!]
\centering \includegraphics[width=1\linewidth]{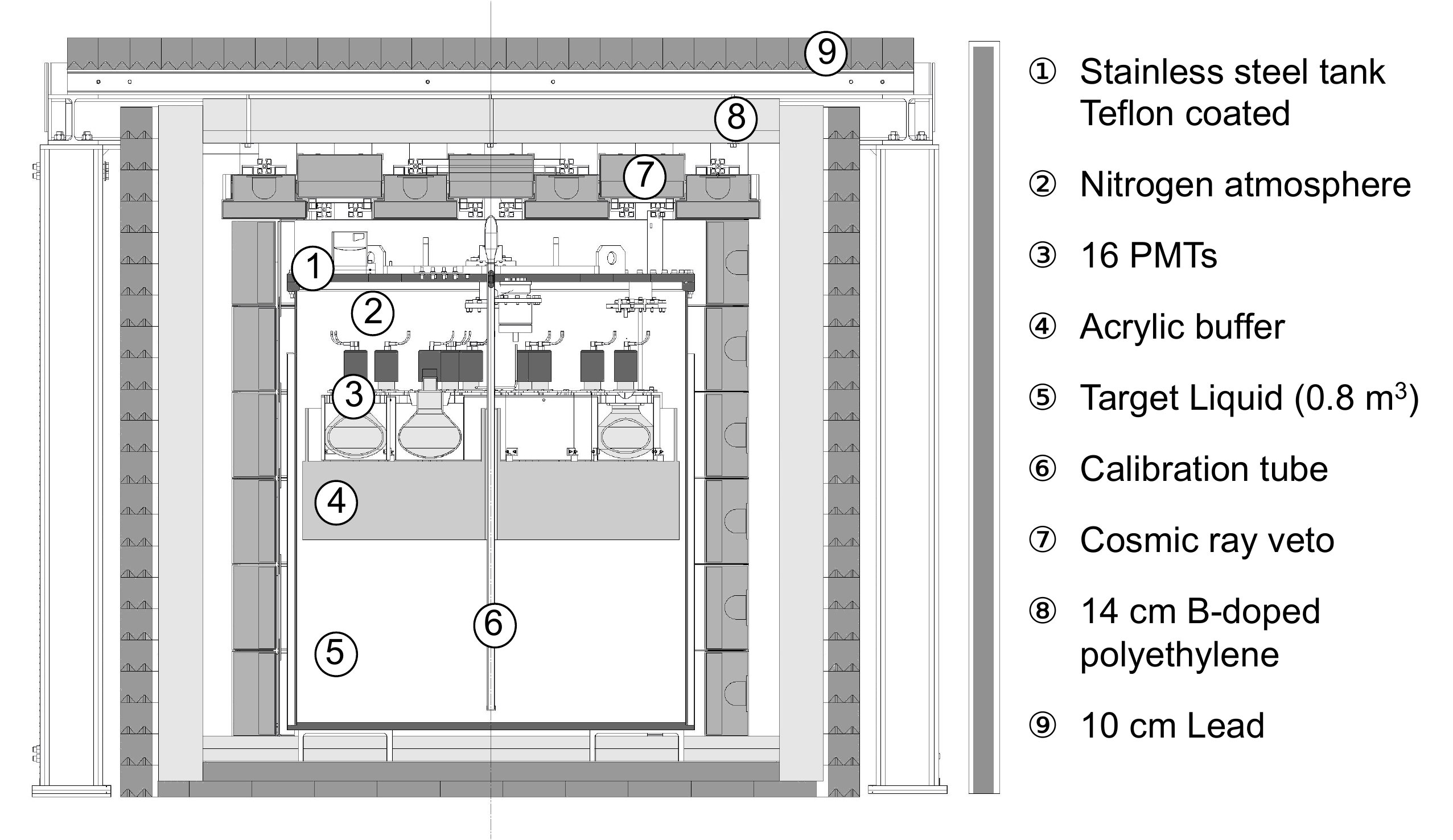}
\caption{Cut view of the Nucifer detector. The overall volume is about \SI{3 x 3 x 2.4}{m}.}
\label{fig:detector}
\end{figure}

The Nucifer detector (Fig.~\ref{fig:detector}) consists of a cylindrical liquid scintillator
tank, surrounded by an active plastic scintillator veto to tag cosmic ray muons, and two layers of shielding:
\SI{14}{cm} of boron-doped polyethylene to capture neutrons, and \SI{10}{cm} of lead layer to attenuate external gamma rays.

Three additional lead walls have been erected to further attenuate the reactor induced gamma rays.
A first \SI{10}{cm} thick wall was originally installed between the detector and the reactor wall,
on the east direction side (see Fig.~\ref{fig:exp_layout}).
Then a second \SI{4}{cm} thick wall was later built between the original lead wall and the reactor wall to further
reduce the amount of gamma rays originating directly from the core (see section~\ref{sec:acc}) resulting
in a factor of 3~additional attenuation.

Finally, a third \SI{10}{cm} thick wall was erected on the southern side of the detector (see Fig.~\ref{fig:exp_layout})
to suppress the gamma radiations coming from the primary water loop of the reactor cooling circuit located behind
the \SI{1}{m} thick concrete wall of the detector casemate.
Indeed when flowing through the fuel elements, the water of this circuit is being highly activated by the fast
neutron flux inside the core. \nuc{N}{16} is produced through $(n,p)$ reaction on \nuc{O}{16} of
the water~\cite{NNDC} and subsequently decay with a \SI{7.13}{s} half-life emitting a \SI{6.1}{MeV}
$\upgamma$-ray with an intensity of \SI{67}{\%} (plus a \SI{7.1}{MeV} $\upgamma$-ray with intensity \SI{4.9}{\%}).
A dedicated circuit, located partly behind the southern wall of the Nucifer casemate,
delays its arrival to the primary pumps by about \SI{90}{s}, giving time for \nuc{N}{16} to decay completely.
The effect of this added shielding was to reduce the single event rate by more than an order of magnitude.

The liquid scintillator is contained inside a stainless steel cylindrical vessel (\SI{1404.2}{mm} in height,
\SI{1250.4}{mm} in diameter). The internal surface of this vessel is coated with reflective white Teflon,
chosen for its chemical compatibility with the liquid scintillator as well as for increasing visible light
collection. The tank contains \Vol{} of liquid scintillator doped with a Gadolinium complex
(a Gd-beta-diketonate) to enhance the capture of thermal neutrons and sign the neutron capture.

The chemistry of the Nucifer scintillator is based on the target liquid of the Double Chooz experiment~\cite{Aberle:2011ar},
with three main modifications.
First the Gd-concentration was increased to \SI{0.17}{\%} in mass to reduce the capture time of thermal neutrons ($\tau_n$),
expected to be in the range of \SI{20}{\us}.
Second the concentration of o-PXE (ortho-Phenylxylylethane) in the scintillator was increased to \SI{57}{\%} in volume,
the remaining \SI{43}{\%} being dodecane. In this way light yield and pulse shape discrimination power for rejection
of correlated background events are improved.
Finally, the concentration of the primary fluor PPO (2,5-Diphenyloxazole) was slightly increased to compensate
for light losses due to the higher Gd concentration~\cite{Aberle:2011ar}. However, the concentration of the
secondary fluor bis-MSB (1,4-Bis(2-MethylStyryl)Benzene) was kept at \SI{20}{mg/l}.

The light collection is performed by sixteen 8-inch photomultipliers (PMTs), type Hammamatsu R5912, located at the top
of the detector vessel. A \SI{25}{cm} thick acrylic disk separates the PMTs from the target liquid scintillator.
Filled with mineral oil, this so-called buffer optically couples the PMTs to the liquid scintillator.
This design allows a good uniformity of the detector response to energy deposition in the whole target volume
and shields the scintillator from the intrinsic PMT radioactivity. The buffer is fixed to the tank lid.
No phenomenon of light emission from PMTs was detected (so-called ``flashing PMTs'').

A light injection system guiding light from LED to Teflon diffusers in the tank through optical fibres
allows the monitoring of the PMT gain as well as the liquid optical properties,
thanks to several types of light diffusers. 
In addition small encapsulated radioactive sources can be deployed along the target central axis inside a
vertical stainless steel tube externally coated with Teflon.

The muon veto is made of 32~modular detectors, each one containing a \SI{5}{cm} thick plastic scintillator bar
(150~to~\SI{170}{cm} length, \SI{25}{cm} width) associated to a single PMT decoupled from its surface. The
plastic scintillator thickness and light collection were optimized in order to discriminate cosmic muons from
high energy gamma rays in a compact and cost-effective way. Indeed, muon energy deposition is at least \SI{10}{MeV},
corresponding to perpendicularly crossing muons at the minimum ionizing point, while the highest gamma energy
is also \SI{10}{MeV}, from neutron capture on metals surrounding the detector or in the concrete.
Using events that saturate all PMTs in the Nucifer vessel and checking if the muon veto had triggered or not,
we measured a muon detection efficiency of \SI{97}{\%}.

\subsection{Electronics and data acquisition}
The analogical output of each of the 16~PMTs is split in 4~different channels. The first channel is used
to build the trigger (the analogical sum of all PMTs), and the second channel is routed to a constant fraction
discriminator to get the time information. The two last channels are delayed and sent to commercial CAEN
Charge to Digital Converter (QDC) modules. One integration gate is adjusted to integrate the full signal
centred around a \SI{150}{ns} time window. The other gate is delayed by about \SI{40}{ns} to integrate only
the late component of the signal. The comparison of the resulting two charges can be later used for Pulse
Shape Discrimination (PSD).

The acquisition is triggered either by the analogical sum of all PMTs overcoming a threshold equivalent
to about \SI{1}{MeV}, or by any of the muon veto module overcoming a threshold equivalent to about \SI{10}{MeV},
or by computer driven LED and random signals.
The dead time is computed online by counting internal clock pulses inhibited by all hardware vetos (such as QDC
busy gates or saturation of the buffers of the QDC/TDC).

The Data Acquisition system (DAQ) is based on the LabVIEW software allowing a remote control of the
acquisition and a constant monitoring of the safety parameters, such as pressure, liquid level and temperature
at various locations. The data stream is automatically chopped up in \SI{500}{MB} runs and transferred to the
CC-IN2P3 computing centre for storage and off-line analysis. Typical run duration is \SI{43}{minutes} with the
Osiris reactor operating at full power, and \SI{53}{minutes} when the reactor is not operating. The difference
comes from the reactor induced gamma background in the Nucifer tank.

In case of a significant deviation of any set of predefined parameters, warning emails are automatically sent
 to on-call experts. If the deviation exceeds a higher threshold, an alarm is sent to the reactor control room.
 Since its installation at Osiris in Spring~2012, the Nucifer detector has been operating without any safety failure.

\section{Expected signal}
\label{sec:Signal}

\subsection{Theoretical neutrino rate}
\label{sec:rate_theo}

The detection of Osiris antineutrinos is achieved through the Inverse Beta Decay (IBD) reaction on the liquid
scintillator free protons: $\bar{\nu_e}+p \rightarrow e^+ + n$, taking advantage of the time delayed coincidence
between the positron and neutron signals. The positron detection is the prompt signal, its energy is related to the
neutrino's by: $\Ep{} \sim \text{E}_{\bar{\nu_e}} - \SI{0.782}{MeV}$. Then the neutron is captured with high
efficiency either on Gd or H atoms in the liquid scintillator. Neutron captures occur with a mean
time of about \SI{20}{\us} and lead to the emission of a few $\upgamma$-rays with a total energy of \SI{8.05}{MeV}
in average for Gd natural isotopic composition and thermal neutron capture at ambient temperature.

To first order, the number of neutrino events detected per day depends on the neutrino flux emitted by the
reactor, the baseline, the number of target free protons, and the detector efficiency. The neutrino flux
depends on the reactor thermal power and, since different fissioning isotopes lead to different neutrino
spectra, on the fission fractions (see Fig.~\ref{fig:FissionFractions}). Finally, the event rate at a given time
$t$ is given by
\begin{equation}
\label{eq:tau_all}
\tau_{\bar{\nu}}(t) = \dfrac{P_\mathrm{th}(t)}{\displaystyle{\sum_k \alpha_k(t) {E_k}}} \, \rho_p \sum_k
\alpha_k(t) \int_{0}^{\infty} \sigma_{\bar{\nu_e}}(E) S_k(E) \, \leff \ud E
\end{equation}
The first term describes the number of fissions per unit time with $P_\mathrm{th}(t)$ the thermal power,
$\alpha_k(t)$ the fission fraction of isotope $k$ ($k=$\Urfive{}, \Ureight{}, \Punine{}, \Puone{}) and
$E_k$ the mean energy per fission of isotope $k$. $\rho_p$ is the proton density in the liquid scintillator.
The integral on the antineutrino energy represents the mean fission cross-section with
$\sigma_{\bar{\nu_e}}(E)$ the IBD cross section and $S_k(E)$ the fission spectrum of isotope $k$ in units of
\si{\neb{}.fission^{-1}.MeV^{-1}}~\cite{Kopeikin:2004cn}.

The effective length $\leff$ is homogeneous to a distance and takes into account the finite
extensions of both the core and the detector, a necessary refinement due to the very short baseline of the
experiment. It is defined as
\begin{equation}
\label{eq:l_eff} \leff = \mathop{\iiint}_{\Vc} \Psi_{\mathrm{f},k}(t,\rc)
\mathop{\iiint}_{\Vd} \dfrac{\varepsilon(E, \rd)} {4 \pi \left( \rd - \rc \right)^2} \ud^3 \rd \ud^3 \rc
\end{equation}
with $\Psi_{\mathrm{f},k}(t,\rc)$ the fission density of isotope $k$, normalized to one fission,
$\Vc$ the reactor core volume, $\Vd$ the detector volume,
$\varepsilon(E, \rd)$ the detection efficiency,
$\rc$ a vector pointing to a given point in the core and
$\rd$ a vector pointing to a given point in the detector.
In the case of point-like core and detector one readily recovers the equivalence $\rho_p  \leff
\equiv \mathcal{N}_p/4 \pi L^2$ with $\mathcal{N}_p$ the number of protons in the target and $L$ the mean
baseline.

\subsection{Parameters and associated uncertainties}
\label{sec:parameters}

All relevant reactor operating parameters are being provided by the Osiris facility, with a \SI{5}{min}
period, leading to an average thermal power of \MeanPower{}.
From the enthalpy balance performed online on the primary circuit of reactor we estimate a \SI{2}{\%} relative
uncertainty~\cite{Pelzer:2012}.

Furthermore the layout and burn-up of each assembly is made available, at the beginning of each reactor cycle.
Detailed core simulations are then performed with the 3D Monte-Carlo code \tripoli{}~\cite{Tripoli:NEA}.
We studied two core configurations, corresponding to the two typical initial burn-up maps of Osiris
with either the control rod number~3 or~4 used as the last rod to control the reactor power.
For each configuration two extreme positions of the control rods, at the beginning and end of the cycle,
have been considered.
During the cycle, the translation of the barycentre of the fissions due to the control rods is only \SI{2.5}{cm},
so the deviation from the average barycentre of the fissions is negligible. 
The corresponding impact on the interaction rate is inferior to \SI{0.3}{\%}.

Between the two initial control rod sequences, the mean barycentre of the fissions is shifted by \SI{3.6}{cm}
at the beginning of a cycle and \SI{1.1}{cm} at the end, leading to a difference in the predicted rate inferior
to \SI{1}{\%}. Still this effect is taken into account for the prediction of the mean neutrino rate.
The overall uncertainty on the fission fraction is assumed to be \SI{2}{\%}, covering the amplitude of the
evolution during a full reactor cycle. Inserted in Eq.\ref{eq:tau_all}, this leads to a \SI{1}{\%} uncertainty
in the predicted neutrino flux. The mean fission fraction are $\alpha_{\Urfive} = \SI{92.6}{\%}$,
$\alpha_{\Punine} = \SI{6.1}{\%}$, $\alpha_{\Ureight} = \SI{0.8}{\%}$ and $\alpha_{\Puone} = \SI{0.5}{\%}$, as expected
for a highly enriched fuel.


The mean energies per fission $E_k$ and the IBD cross-section are taken from~\cite{Kopeikin:2004cn}
and~\cite{Strumia:2003zx} respectively. For the antineutrino fission spectra of each isotope we used
the recently improved spectra from~\cite{Huber:2011wv}, converted from the ILL reference beta spectra of \Urfive{},
\Punine{} and \Puone{}~\cite{Schreckenbach:1985ep,Hahn:1989zr}. The \Ureight{} spectrum is taken from~\cite{Haag:2013raa}.
Considering the dominant contribution from \Urfive{} a global \SI{2.2}{\%} relative uncertainty is obtained
for the mean interaction rate of antineutrinos.

The distance between the reactor core centre to the detector centre was determined using a survey of the geodesic
combined with information from technical drawings of the reactor building. This led to a baseline of \Dist{},
corresponding to a \SI{3.1}{\%} uncertainty on the neutrino rate.
The detector mechanical survey allowed us to compute the detector volume at \Vol{}.
Associated to the measurement of the target liquid mass of \SI{739(1)}{kg} and the calculation
of the ratio of free hydrogen over carbon from the known chemical composition of the liquid (H/C=\num{1.50(2)}),
we computed a proton density of \SI{5.92(5)e28}{proton/m^3}.

To propagate the \neb{} flux from the core to the \neb{} detected rate we developed a code called NuMC,
with main task to integrate numerically Eq.~\ref{eq:tau_all} and~\ref{eq:l_eff} with a Monte-Carlo method.
The code firstly computes the interaction rate for a perfect detector: a given fissile isotope is drawn according to
the simulated fission fraction $\alpha_k$, then a \neb{} energy is drawn in the corresponding spectrum $S_k$, and finally
a position in the core is drawn according to the simulated distribution of fissions in the core $\Psi_{\mathrm{f},k}(\rc)$.
From this stage, the \neb{} is propagated in a random direction, and the length of the \neb{} path inside the detector
(if any) is stored. Repeating a great number of times this procedure gives the average length of the \neb{} path
inside the detector, which is \leff{} weighted by the fission fraction.
After numerical integration of the product of the spectra $S_k$ and the cross-section $\sigma_{\bar{\nu_e}}$,
the interaction rate is then obtained by applying the correct normalization factor of Eq.~\ref{eq:tau_all}.
We obtain for Osiris at \MeanPower{} and \SI{100}{\%} of efficiency \RateNuMCnoEffConfigA{} and \RateNuMCnoEffConfigB{}
for the two core configurations described previously, with a relative uncertainty of \SI{4.6}{\%}
(see table~\ref{tab:flux_sys}).

Then we can compute the expected number of \si{\neb/day} according to:
\begin{equation}
\Rnu{pred} = (\Rrod{3}_\nu \times \Trod{3} + \Rrod{4}_\nu \times \Trod{4}) \times \epsilon_{\text{det}}
\end{equation}
with $\Rrod{3}_\nu=\RateNuMCnoEffConfigA$ ($\Rrod{4}_\nu=\RateNuMCnoEffConfigB$) the expected number of \neb{}
interactions per day at \MeanPower{} with the 3\textsuperscript{rd} (4\textsuperscript{th}) control rod used as
the last control fuel element during the cycle. $\Trod{3}=0.32$ and $\Trod{4}=0.68$ are the relative amounts of lifetime
in each configuration and $\epsilon_{\text{det}}$ the detection efficiency (see section~\ref{sec:DetEff}).
We obtain $\Rnu{pred} = \SI{913}{\neb{}/day}$ for Osiris at \MeanPower{} and \SI{100}{\%} of efficiency.

\begin{table}[!ht]
\begin{center}
  \begin{tabular}{l>{\centering}m{2.5cm}}
    \toprule
      Source 		        & Relative uncertainty (\%) 	\tabularnewline
    \colrule \hline
      Baseline	        	& 3.1	\tabularnewline
      Fission cross-section	& 2.2	\tabularnewline
      Thermal power		& 2.0	\tabularnewline
      Fission fractions	        & 1.0	\tabularnewline
      Number of protons	        & 1.2	\tabularnewline
    \colrule
      Total     		& 4.6	\tabularnewline
    \botrule
  \end{tabular}
\end{center}
\caption{\label{tab:flux_sys} Summary of relative uncertainties associated to the predicted \neb{} rate,
with \SI{100}{\%} detection efficiency.}
\end{table}

In a second step, the NuMC code can be used as a Geant4 IBD event generator. An isotope and a fission vertex in the core
are drawn, together with a vertex in the detector. This couple of vertices is stored or not according to the
$1/L^2$ flux law. If this event is valid, an IBD event is generated according to the double differential IBD
cross-section~\cite{Strumia:2003zx}. Finally the full kinematic information of both the positron and the
neutron are saved in a data file as an input for Geant4.

We then simulate the \neb{} interactions in the detector, including scintillation and light collection on PMTs.
The Geant4 output file is set to the same format as the Nucifer data. Finally processing our neutrino search algorithms
on the simulated data determine the mean detection efficiency, found to be \EffPC{} (see section~\ref{sec:DetEff}).

\section{Detector calibration}
\label{sec:calib}

The calibration system was designed to provide an absolute energy scale to the light response of the detector
and to assess the linearity and the stability of the whole detection system. Controlled light
injections allow a calibration from charge unit to photo-electrons (PE hereafter). Radioactive sources
deployed temporarily in the central tube inside the detector let us to perform the final
calibration from PE-scale to the MeV-scale.

The detector target vessel is equipped with 7~Teflon light diffusers, linked by optical fibres to
Light-Emitting Diodes (LED) located in the electronic rack and controlled through the acquisition software.
One of this LED is used to generate Single Photo-Electrons (SPEs) on the PMTs, 4~are used with different
intensities to test the linearity and stability of the PMTs over their full dynamic range, and the last
2~could be used as spares. Sequences of 14~patterns, including 2~SPEs, random and multiple LED patterns, are
continuously generated by the acquisition at a frequency of \SI{5}{Hz}. What is referred to as a random pattern
corresponds to no LED and, although periodic, is random with respect to physical triggers and therefore samples
the pedestals of the digitization chain. This system allows the continuous measurement of both the pedestal and
gain of each channel for each run, thanks to an automatic fit of the pedestals and of the SPE signals. Each run
and channel is consequently automatically calibrated in photo-electrons, allowing a proper sum of the
16~PMT charges.

\begin{figure}[h!]
\centering
\includegraphics[width=1 \linewidth]{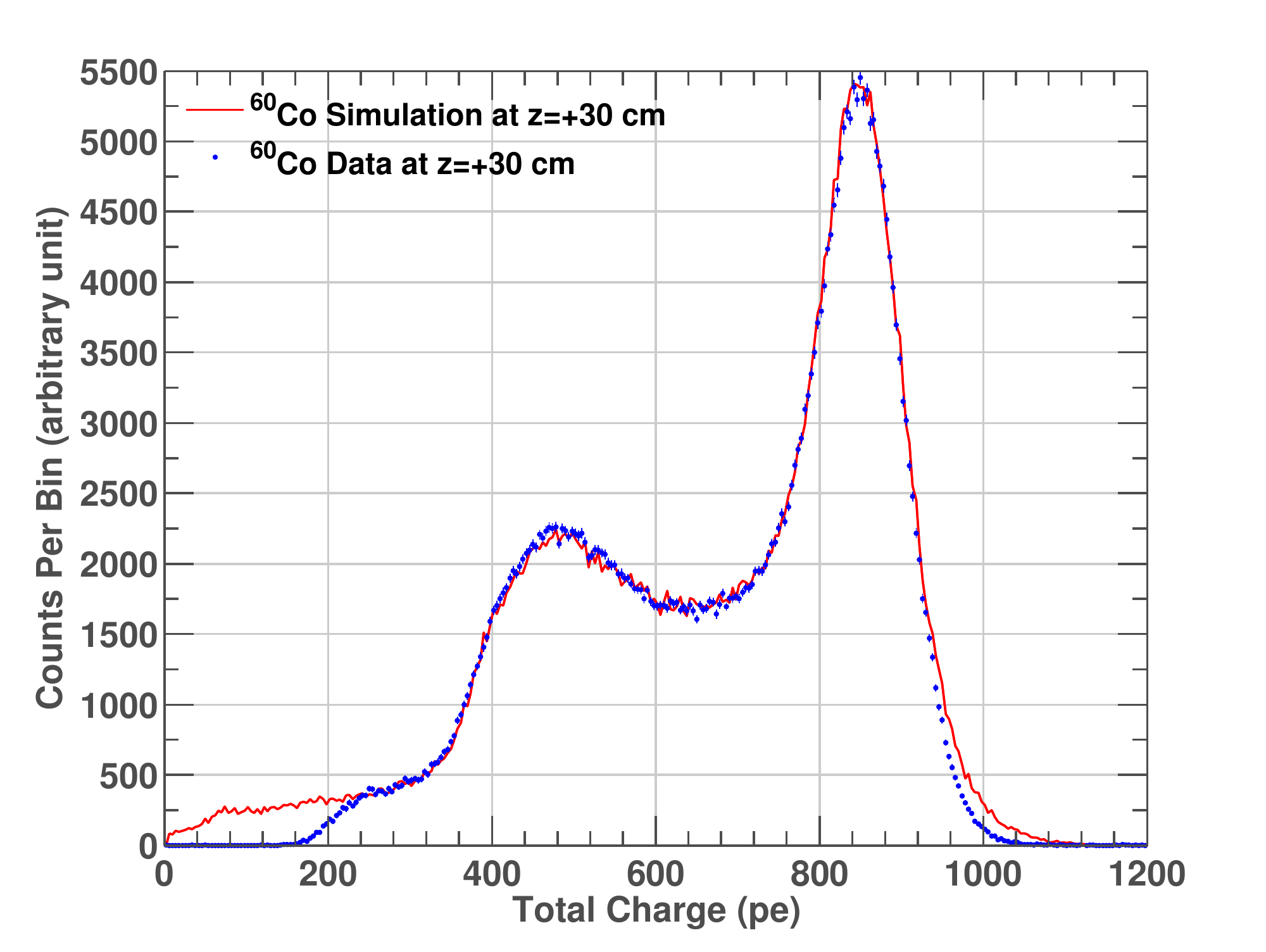}
\caption{Detector response to a \Co{} source deployed at the centre of the target vessel compared to the
detector response simulation. No threshold was applied in the simulation, leading to the low energy discrepancy.
For all calibrations, the hardware threshold is different from the data taking threshold,
adapted to the energy range and the trigger rate.}
\label{fig:60Co_centre}
\end{figure}

The detector response to multiple LED patterns compared to single LED pattern is linear within \SI{1}{\%},
showing no drift at higher intensities and the uncertainty on pedestals is below \SI{1}{\%}. On a run-to-run
basis the variance of the fitted SPE is typically \SI{2.5}{\%} with perfect Gaussian shape. On longer time
scales the gains and the linearity of the response to LED patterns have shown remarkable stability. Larger
drifts due to temperature changes in the electronic rack are observed on the absolute value of pedestals and
LED signals. They are correlated on all channels and corrected for each run.

Small encapsulated radioactive sources can be deployed inside a vertical tube along the target central axis.
Three gamma-emitting sources of few kBq activity have been used: a \Cs{} source with one \SI{661}{keV}
$\upgamma$-ray, a \Co{} source with two $\upgamma$-rays of 1173 and \SI{1332}{keV} in coincidence, and a \Na{}
source with one $\upgamma$-ray of \SI{1274}{keV} in coincidence with two $\upgamma$-rays of \SI{511}{keV}
coming from the annihilation of the positron. We also used a neutron source of \nuc{Am}{241}-Be, emitting a neutron
in coincidence with a $\upgamma$-ray of \SI{4.4}{MeV} for $\sim \SI{75}{\%}$ of the events. The activity
of the \nuc{Am}{241} reaches few MBq, leading to only $\sim \SI{30}{Hz}$ of neutrons.
This source is mainly used to test our analysis procedure for searching correlated pairs of events
and study the neutron physics (capture time and detection efficiency).

These calibration sources were inserted at different elevation levels in the central vertical tube and used to tune
a Geant4.9.4-based simulation (in PE units) to reproduce the measurements. A good agreement between experimental
and simulated spectra could be achieved for most positions (see Fig.~\ref{fig:60Co_centre}). However at the
highest source location, close to the acrylic buffer coupling to the PMTs, sizeable deviations from the data
still remain, likely due to the difficulty of properly simulating light collection for events interacting close
to the PMTs.

\begin{figure}[h!]
\centering
\includegraphics[width=1 \linewidth]{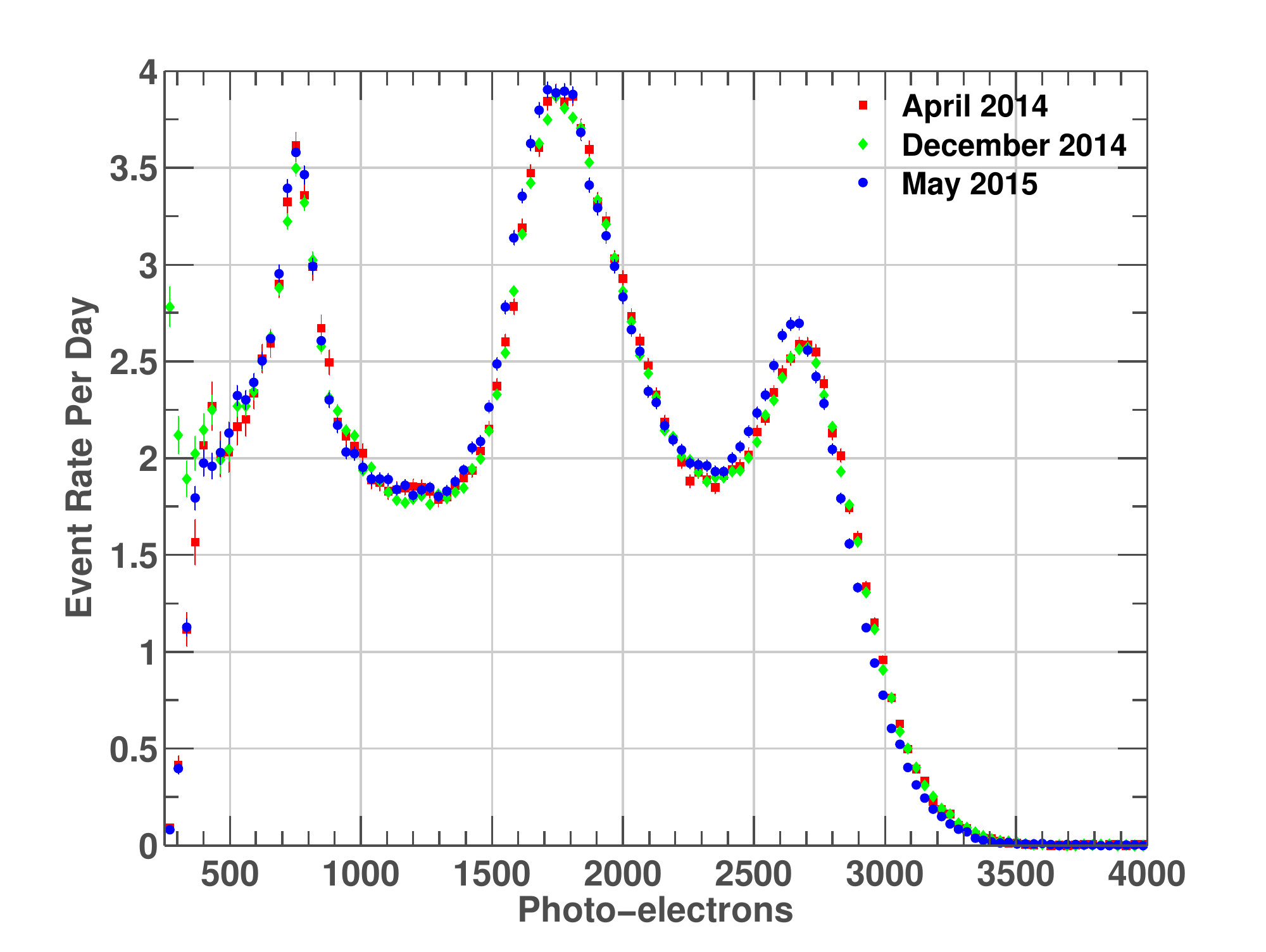}
\caption{Detector response to a AmBe source deployed at the centre of the target vessel at three different
periods from April 2014 to May 2015.
The right peak corresponds to the neutron capture on Gd at \SI{8}{MeV}. The middle peak comes from the reaction
of $\upalpha$ particle on \nuc{Be}{9}, leading to the emission of a \SI{4.4}{MeV} gamma ray and a fast neutron,
and depositing a mean energy equivalent to an electron of \SI{5.5}{MeV}. Due to the quenching effect, neutron induced
nucleus recoils do not produce as much light as an electron of the same energy.
The left peak corresponds to neutron capture on hydrogen, with the emission of a unique \SI{2.2}{MeV} gamma ray.
}
\label{fig:AmBe_centre}
\end{figure}

A global calibration factor of $\sim \SI{340}{PE/MeV}$ has been obtained with the 4~sources of calibration.
The uncertainty on this factor is latter included in the global energy scale uncertainty (see section~\ref{sec:Systematics}).
With a source at the centre of the target we measure an intrinsic energy resolution of \SI{10}{\%} at \SI{1}{MeV}.
This value increases to about \SI{20}{\%} for vertices uniformly distributed in the target volume because of the light
absorption on the Teflon-coated walls of detector target. Figure~\ref{fig:AmBe_centre} illustrates the very good stability
of the detector response to the AmBe source along the whole data taking period.


\section{Data analysis}
\subsection{Data sample}
\label{sec:DataSample}

The signature of IBD events consists in a delayed coincidence starting by a prompt positron energy
deposition, \Ep{}, followed by a neutron induced energy deposition, \Ed{}, due to the de-excitation gamma
ray(s) after its capture on H or Gd within \Delrel{}.
The principle of the analysis is to compare the number of \neb{}
detected to the prediction based on the reactor data and reactor core simulations. The challenge consists in
the statistical separation of IBD events with respect to backgrounds induced either by random coincidences of
single events or by correlated coincidences originating from air showers and possibly from reactor core radiations.

At Osiris, a running cycle is nominally operating for three weeks with the reactor ON followed by a week of
reactor OFF period. The results described in this paper are based on 10 cycles accumulated from June 2014 to
July 2015, after the Nucifer upgrades.
Power transients at the beginning of each cycle (few hours) are discarded from the data sample due to the
difficulty to assess the thermal power of the reactor during this phase. Therefore our sample consists of
\SI{200}{days} of reactor ON at full power. Nevertheless the lifetime of the Nucifer data acquisition is only \SI{145}{days}.
A loss of \SI{5}{days} is due to the \SI{100}{\us} veto of the acquisition after each tagged muons. The
remaining inefficiency is dominated by unattended periods of data taking during which the synchronization of
the different electronics readout buffers of the data acquisition system was lost. The bad runs are rejected
by an automatic off-line quality check. The total of reactor OFF time is \SI{153}{days} including inter-cycles
periods and longer shutdowns for reactor maintenance. The Nucifer lifetime during this OFF period is
\SI{106}{days}. We thus obtain an overall data taking efficiency of \SI{70}{\%}. It is worth noting that this
efficiency could easily be improved to more than \SI{90}{\%} by implementing an automatic recovery procedure
to handle data acquisition system failures.

In the following, event rates are reported with statistical uncertainties only. Systematic uncertainties are gathered
in the predicted neutrino rate uncertainty.

\subsection{Neutrino candidate selection}
\label{sec:NuCuts}
The analysis cuts are optimized to reach the smallest relative uncertainty on the detected neutrino rate~\cite{Nghiem:2013}
in the severe background conditions described in the sections~\ref{sec:acc} and~\ref{sec:cosmic_bkg_ori}.

From the prediction of the emitted neutrino spectra and the kinematics of the IBD reaction the expected range
of \Ep{} is \SI{1.022}{MeV} to about \SI{7}{MeV}. Given the steep increase of background at low energy, a \SI{2}{MeV}
threshold is used, well above the hardware threshold, giving negligible trigger inefficiency.
Then the maximum prompt energy is set at \SI{7.1}{MeV} above which the remaining IBD positron induced rate is negligible.

The delayed event energy is not related to the incident neutrino energy. Its associated \SI{8}{MeV}
$\upgamma$-cascade from the neutron capture on a Gd nucleus is used to efficiently discriminate the neutrino
signal against lower-energy background events. However in a meter-scale detector like Nucifer one has to
accommodate for important energy leakages shifting most of the delayed events into lower energy
bins. Indeed, the most probable energy deposited in Nucifer after a neutron capture on Gd is about \SI{4.5}{MeV} only.
Again the limitation to open the \Ed{} range towards lower energies is the steep increase of the
accidental background. We thus optimized the delayed energy range to $\SI{4.2}{MeV}<\Ed{}<\SI{9.6}{MeV}$.

The coincidence time window \Delrel{} is limited to \SI{40}{\us} after each prompt candidate. This duration
corresponds to only twice the expected neutron capture mean parameter, $\tau_n$. 
This upper limit is a compromise between the rejection of about \SI{15}{\%} of the neutrino signal
and the mitigation of the accidental background rate, growing linearly with the prompt-delayed gate duration.
Furthermore the delayed event time must also be separated from the prompt energy deposition by more than \SI{6}{\us}
due to the dead time of the QDC needed to process the PMT signals after each trigger.

\begin{figure}[h!]
\centering \includegraphics[width=1 \linewidth]{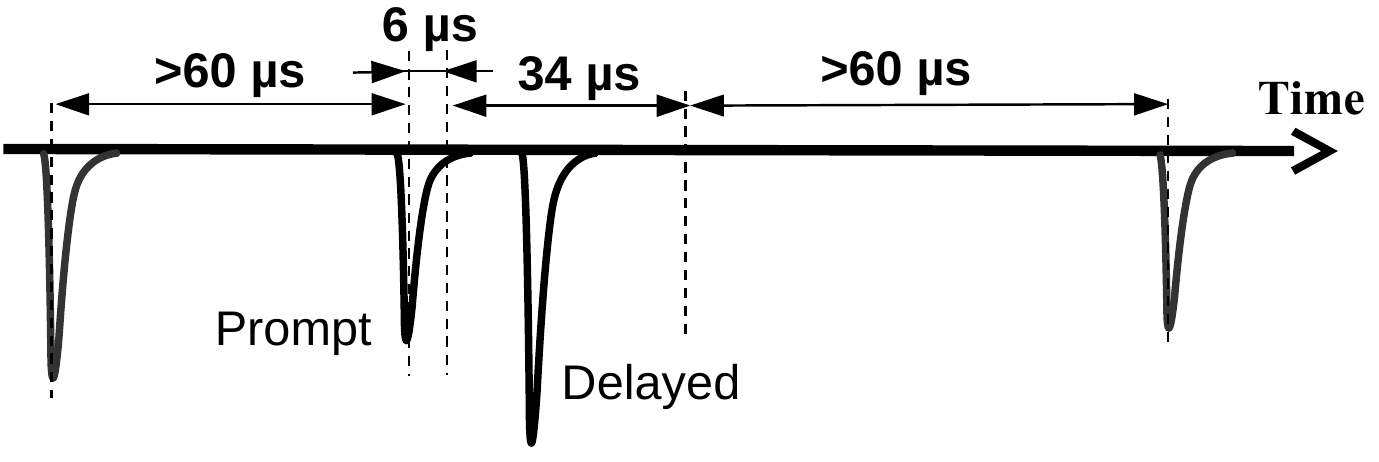}
\caption{\label{fig:pulse_corr} Time selection of neutrino candidates. The prompt event must be separated by
more than \SI{100}{\us} with respect to any muon veto trigger.}
\end{figure}

All events occurring less than \SI{100}{\us} after a muon event are overridden. A muon event is defined either
as a muon veto trigger or as the saturation of at least 15~PMTs.
Finally an isolation selection criteria, called multiplicity cut, is applied to all prompt-delayed
pairs. It imposes that no energy deposition occurs either \SI{60}{\us} before the prompt event or
\SI{60}{\us} after the end of the delayed gate (see Fig.~\ref{fig:pulse_corr}). This cut rejects cosmic-ray-induced
background with more than two particles in the same shower. All selection criteria are summarized in
table~\ref{tab:SelectCuts}.

\begin{table}[!ht]
\begin{center}
    \begin{tabular}{l @{~~}l @{}m{0pt}@{}}
	\toprule
	Criterion	& Applied cut					&\tabularnewline \colrule
	Prompt event	& $\SI{2.0}{MeV}<\Ep{}<\SI{7.1}{MeV}$		&\tabularnewline[2pt]
	Delayed event	& $\SI{4.2}{MeV}<\Ed{}<\SI{9.6}{MeV}$		&\tabularnewline
	Time selection	& $\SI{6}{\us}<\Delrel{}<\SI{40}{\us}$		&\tabularnewline
	Multiplicity	& No trigger \SI{60}{\us} before prompt event	&\tabularnewline
			& No trigger \SI{60}{\us} after delayed gate	&\tabularnewline
	Muon veto	& \SI{100}{\us}					&\tabularnewline
 	\botrule
 \end{tabular}
  \end{center}
\caption{\label{tab:SelectCuts} Summary of neutrino selection criteria.}
 \end{table}

In the following we will focus on the accidental and the correlated backgrounds. Indeed the very
short baseline coupled with the shallow depth of the experiment imply that these backgrounds make
much larger contributions than the expected neutrino signal does. Therefore an accurate
determination of these spurious event rates is the main challenge of the Nucifer analysis.

\subsection{Accidental background}
\label{sec:acc}
\begin{figure}[h!]
\centering \includegraphics[width=1 \linewidth]{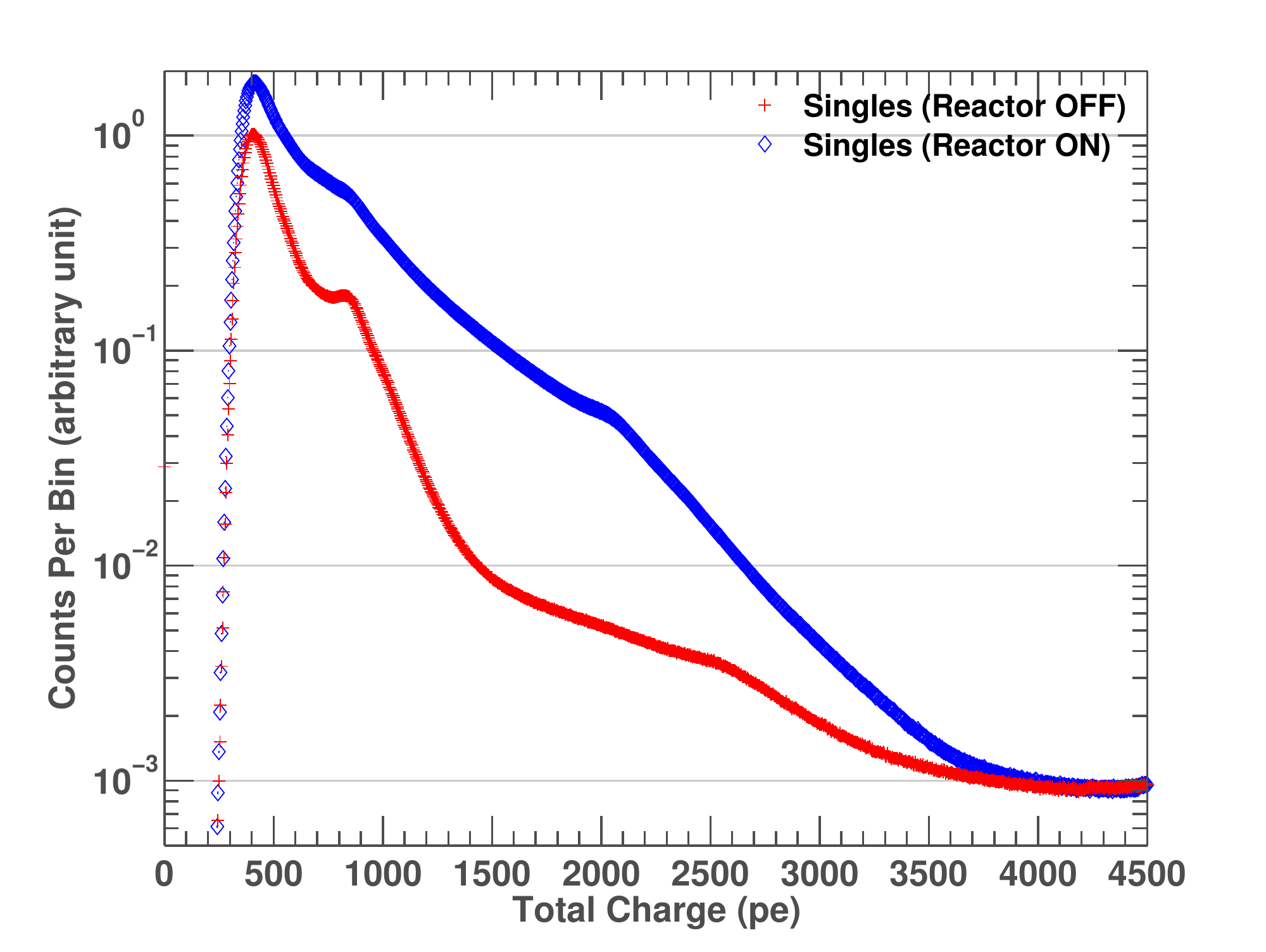}
\caption{\label{fig:spectra_single} Energy spectrum of the single event rate at reactor full power
and during reactor OFF combined periods.}
\end{figure}

When the reactor is OFF, the trigger rate is dominated by muons events ($\sim \SI{350}{Hz}$ in the muon veto),
and once muons have been excluded, by low energy natural radioactivity decays (\SI{65.7}{Hz} above \SI{2}{MeV})
whose energy spectrum is shown in red on figure~\ref{fig:spectra_single}.
Beyond the rapid decrease of the energy spectrum after the threshold at about \SI{500}{pe}, one can see a small peak
at \SI{2.6}{MeV} (\SI{880}{pe}) corresponding to the highest natural gamma ray of \nuc{Tl}{208} decay
(\Co{} in the tank stainless steel is less than \SI{12}{mBq/kg} from a sample measurement and therefore negligible),
and an ankle near \SI{8}{MeV} (\SI{2700}{pe}) attributed to natural neutron capture.
The tail at high energy is attributed to the contamination by events induced by atmospheric showers
in the vicinity to the detector that do not trigger the muon veto nor saturate at least 15~PMTs.

When the reactor is ON the shape is different and the total event rate increase (\SI{177.1}{Hz} above \SI{2}{MeV}).
The associated spectrum (blue curve on figure~\ref{fig:spectra_single}) shows a rather smooth shape up to
about \SI{10}{MeV}. Such high-energy depositions are attributed to neutron radiative captures on metallic structures
and concrete wall components, such as iron, nickel or aluminium, occurring in the vicinity of the detector
and emitting high-energy $\upgamma$ that can pass through the shielding and finally reach the detector target.
The bump around \SI{6}{MeV} (\SI{2000}{pe}) comes from the \nuc{N}{16} decay gamma rays in the deactivation circuit
(see section~\ref{sec:Detector}).

As a consequence both the prompt and the delayed energy windows are affected by the reactor induced events, which has
a strong consequence: the accidental background event rate scales quadratically (instead of linearly)
with the gamma-ray flux $\phi_\upgamma$, proportional to the reactor power.
A crude model of the neutrino-like event rate is $\taunu=\taupr\times\taudel\times\Delrel$. Both prompt and delayed
event rates are dominated by backgrounds events, and if both are affected by reactor induced events we get
$\taupr\propto \phi_\upgamma$, $\taudel\propto \phi_\upgamma$ and finally $\taunu\propto \phi_\upgamma^2$.
Adding lead shielding could therefore reduce greatly the accidental background, but the mass limit on the Osiris floor
has already been reached.

\begin{figure}[h!]
\centering \includegraphics[width=1 \linewidth]{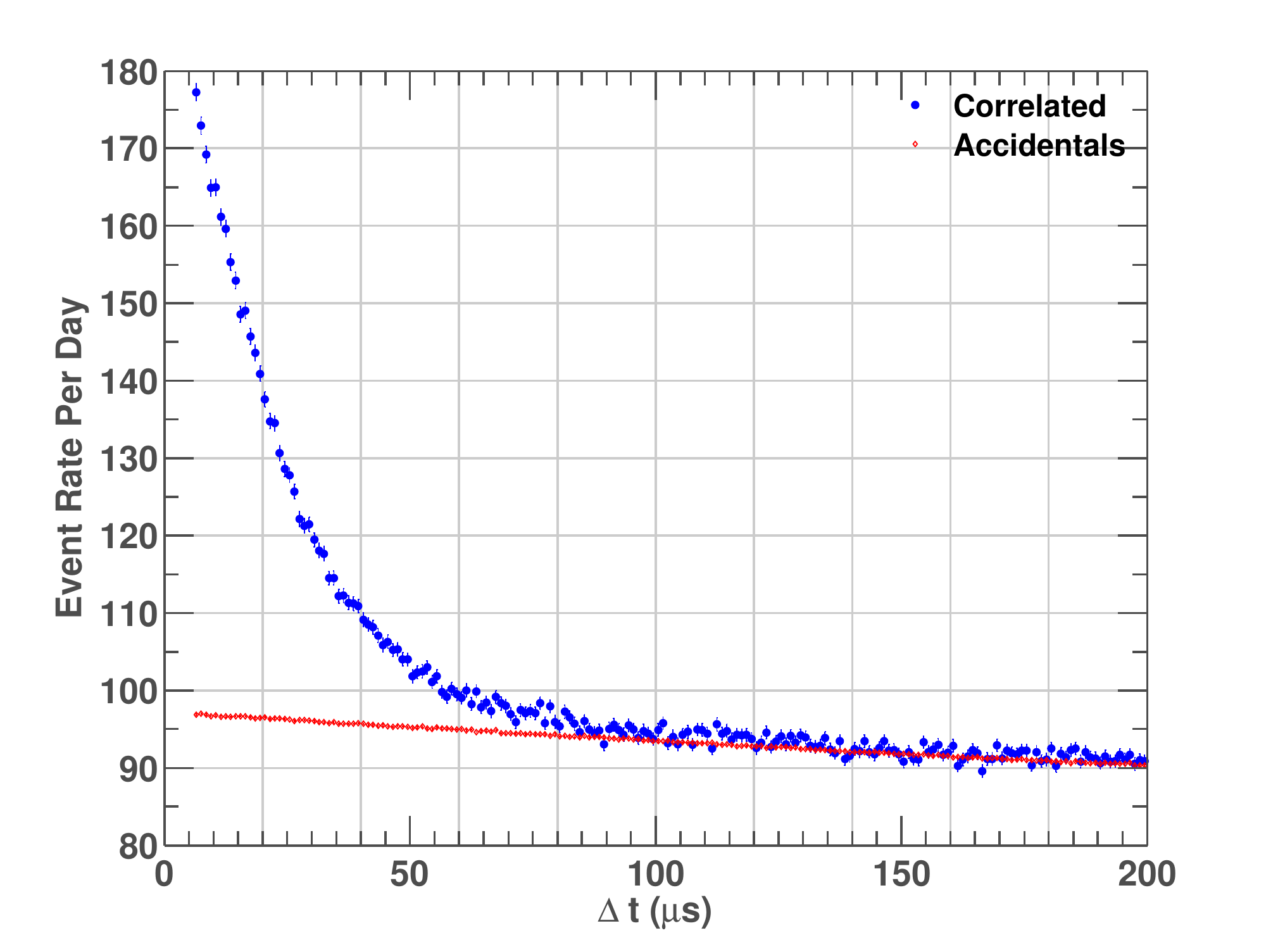}
\caption{\label{fig:Deltat_ON} Distribution of the time difference between prompt and delayed events for all
correlated candidates (blue) and for the accidentals (red). The data sample accumulates 10~reactor cycles.}
\end{figure} 

Nevertheless this high accidental background level can be accurately measured by analysing off-time coincidences.
For each delayed-like event, all past prompt events in the run are shifted in time up to 100~times
by steps of \SI{1}{ms}, until the next past prompt is repositioned less than \SI{1}{ms}
before the current delayed event. If the so-formed virtual pair complies with the neutrino selection cuts, it
is counted as an accidental event. This method allows to measure the accidental background 100~times for each
run, in the exact same conditions of data taking, and consequently this background is measured with a negligible
statistical uncertainty. Unfortunately, this method limits the impact of the high accidental event rate on the final
statistical uncertainty only for one of the two terms of the subtraction (see also table~\ref{tab:rates}). Noting $N$
the event number, $\Delta t$ the acquisition time, and $\tau$ the events rates, the correlated event rate error is:
\begin{align}
 \err{corr} &= \sqrt{ \left(\frac{\sqrt{\Num{candidate}}}{\Delta t}\right)^2  + \left(\frac{\sqrt{\Num{acc}}}{100 \times \Delta t}\right)^2 } \\
 &= \sqrt{ \frac{\taug{corr}+\taug{acc}}{\Delta t}  + \frac{1}{100}\frac{\taug{acc}}{\Delta t} } \sim \sqrt{ \frac{\taug{acc}}{\Delta t} }
\end{align}
where we see that if the 100~shifted gates allow to neglect the uncertainty brought by the accidental events subtraction,
the final uncertainty is still dominated by the accidental event rate, as $\taug{corr}\ll\taug{acc}$.

Furthermore, great care is brought into the evaluation of the efficiencies of the multiplicity cut and muon veto
as they slightly differ between the searches for correlated pairs and accidental pairs. A cross-check of our accurate
determination of accidental rate is illustrated in figure~\ref{fig:Deltat_ON} where the number of measured correlated
and accidental pairs is plotted as a function of \Delrel{} for a same sample of reactor ON data. The correlated events
clearly show up as an exponential curve on top of the accidental background with a decay time corresponding
to the n-capture time $\tau_n\simeq \SI{20}{\us}$. The distribution of pure accidentals, determined with the off-time
prompt events, perfectly matches the correlated curve for $\Delrel\gg\tau_n$.

Within the selection cuts, summarized in table~\ref{tab:SelectCuts}, the average accidental rates
measured for the whole data sample are \RateAccOFF{} when the reactor is OFF and \RateAccON{} when the reactor is ON
(at full power), respectively. Therefore this dominating background is \AccToSignalRatio{}~times larger than
the expected neutrino rate. In the following all quoted rates will be corrected for their associated accidental rates.

\subsection{Cosmic induced background}
\label{sec:cosmic_bkg}
\subsubsection{Origin and rejection}
\label{sec:cosmic_bkg_ori}
The correlated events above the accidentals events in figure~\ref{fig:Deltat_ON} are still not a pure sample of neutrinos.
At the shallow depth of Nucifer (\SI{12}{mwe}), the cosmic-ray particles induce correlated backgrounds
through the multiple secondary particles produced in the air shower and also through their interaction in the ceilings
above the detector (especially for the penetrating muons).
It is worth noting that the overburden above the detector is not sufficient to stop all the hadronic component
of the atmospheric cascades, leaving the possibility for some fast neutrons to reach the liquid scintillator.

The most probable candidates at the origin of cosmic induced correlated backgrounds are fast neutrons
from a air shower or created by an inelastic muon interaction (spallation) in materials above or nearby the detector.
First, a fast neutron can scatter off nuclei in the detector target, mimicking prompt-like energy deposition,
and later be captured on Gd providing a delayed-like energy deposition with a similar prompt-delayed
time correlation than expected for neutrinos. Second, two neutrons from the same shower can be captured successively
in the detector, after some diffusion in the liquid or in the shielding. As the gamma collection efficiency
after neutron capture on Gd is poor in Nucifer, the first neutron capture energy deposit is very likely to be in the prompt
energy window. Third, fast neutron can also produce high energy gamma rays by inelastic scattering on nuclei, such as
the $\nuc{C}{12}(n,n')\upgamma$ reaction producing \SI{4.4}{MeV} gamma rays on the first nuclear excitation level.
The gamma interaction would of course mimic the prompt event and the later neutron capture the delayed event.

Fortunately, this last background process whose signature is very similar to the IBD can be safely neglected with respect
to the first process (elastic scattering mimicking the prompt event): the $\nuc{C}{12}(n,n')\upgamma$ cross-section
is always well below both the hydrogen and the carbon elastic scattering cross-section, by a factor 5 to 10 each
depending on the energy~\cite{NNDC}.

Highly energetic muons may also create long-lived $\upbeta$-n emitters such as \Li{} or \He{} when interacting
with Carbon nuclei belonging to the scintillator molecules. The rate of $\upbeta$-n events from \Li{} or \He{}
decays was estimated to \SI{2.7}{events/day} by scaling the measurement of the Double Chooz experiment to the
Nucifer shallow depth~\cite{Abe:2012ar}. We compared this estimation with our analysis of events following
high energetic muon depositions in the detector target at time differences compatible with the decay time of
\Li{} ($\sim \SI{270}{ms}$). The muons showering inside the Nucifer target could be tagged by lowering the
gain of one target PMT. No \Li{} induced background candidate was clearly identified. In consequence an upper
limit of less than \SI{12}{events/day} at \SI{95}{\%}~C.L. was set~\cite{Nghiem:2013}. Therefore this
background can be considered as negligible with respect to the total rate of correlated events.

As the muons entering the detector are tagged by the veto system surrounding the detector, most of the secondary products
of the muon interactions are removed from the data sample by discarding events occurring within \SI{0.1}{ms}
after each recorded muon. A further background reduction is provided by the multiplicity cut, rejecting most of
the neutron induced background. {\it In fine}, the remaining contribution is subtracted using reactor-OFF data,
representing about \SI{45}{\%} of the data taking time. Averaged over the whole reactor OFF data, a correlated event
rate of \RateCorrOFF{} is measured, i.e.~\CorrToSignalRatio{}~times higher than the expected neutrino rate.

\subsubsection{Stability and correction}
\label{sec:cosmic_bkg_stab}
Last but not least, before subtracting this correlated background to obtain the experimental neutrino rate,
one has to assess its stability over the time scale of a reactor cycle. Indeed the correlated background is
found to vary over time and consequently an extra probe of the correlated background must be defined to
monitor and subtract properly the actual background rate during each reactor ON period.

\begin{figure}[h!]
\centering \includegraphics[width=1 \linewidth]{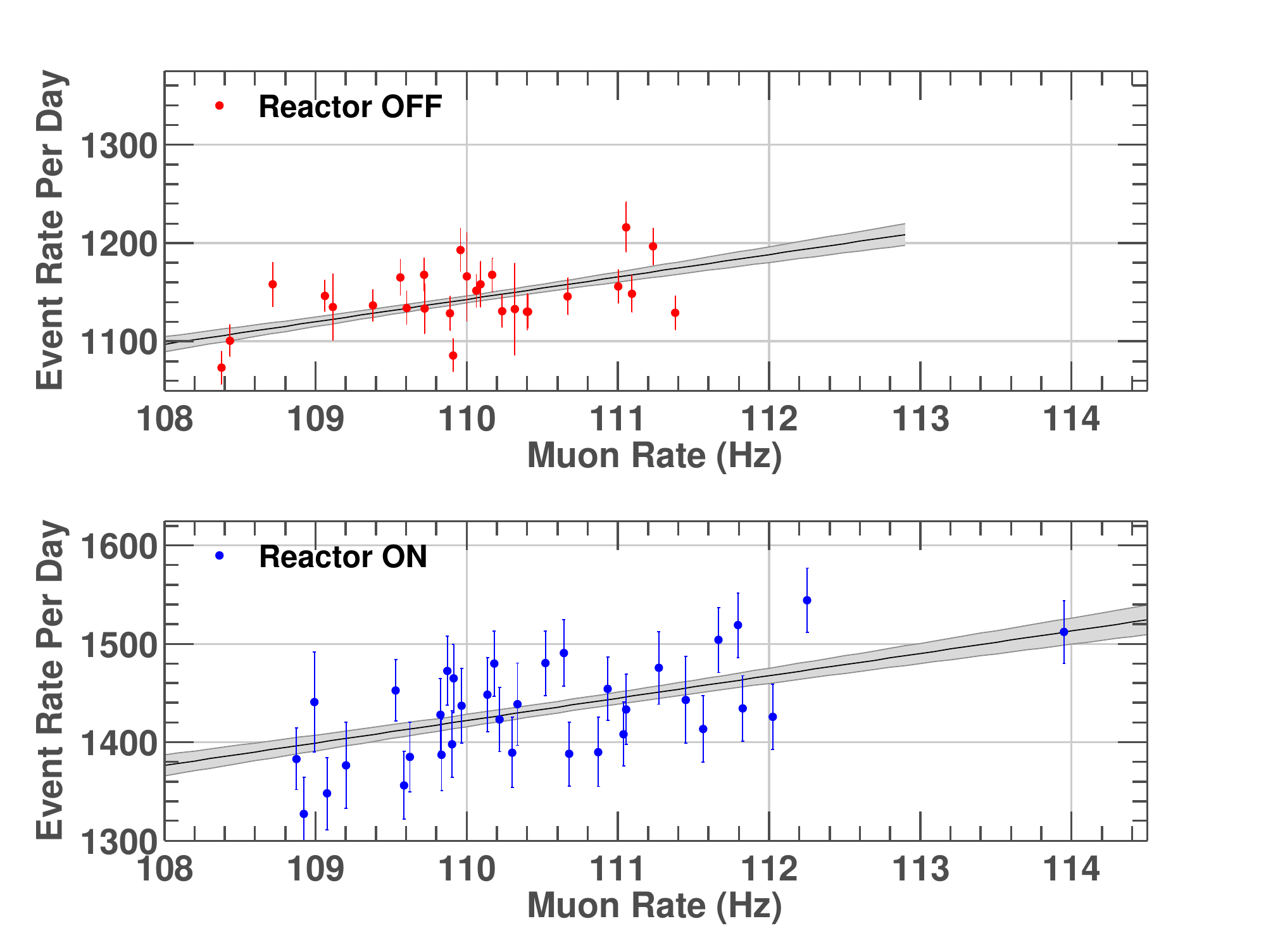}
\caption{\label{fig:Muon_ONOFF} Correlation between the rate of correlated background measured with the
reactor ON/OFF and the rate of triggers tagged as muons. The slope is \SI{-22.8(50)}{event.day^{-1}.Hz^{-1}}.}
\end{figure}

A probe is the muon rate, accurately recorded for each run. Here we define a muon trigger as the
combination of a signal in the muon veto and the saturation of at least 15~PMTs. This definition
reduces the sensitivity to high-energy $\upgamma$ that can trigger the muon veto when the reactor
is ON and allows a reliable comparison between reactor ON/OFF muon rates. Table~\ref{tab:rates} shows that
indeed the muon rates measured this way for the averaged reactor ON and OFF periods are very close,
\num{110.6} and \SI{110.1}{Hz} respectively. The main origin of this variation of muon rates is the natural variation
of the atmospheric pressure: we could check that the remaining \SI{0.5}{Hz} difference is coherent
with the different mean atmospheric pressures recorded for the two averaged periods (998 and \SI{1002}{hPa})
and the observed correlation of \SI{-0.117(8)}{Hz/hPa}. Figure~\ref{fig:Muon_ONOFF} (top)
shows that the variations of the correlated event rate among various reactor OFF periods are very well
correlated with the muon rate. Once the correlation with pressure is removed no sizeable dependence on temperature
or humidity was found.

At first glance one would expect a zero intercept and a relative change of the muon rate equal to the relative
change of the correlated background. This is clearly not observed. More complex physics must be brought into
play to explain the true dependence, probably due to the fact that the tagged muons saturating the detector
are not fully representative of those inducing the correlated background (coming from interactions outside the
detector). Hence a linear fit is considered more as an effective model within the limited range of variation
of the muon rate. What matters for the subtraction is that the same correlation applies on the reactor-OFF
data. This is indeed the case as shown in figure~\ref{fig:Muon_ONOFF} (bottom) where the muon rate dependence
measured for reactor OFF periods is in very good agreement with the one for reactor ON data. Hence the
subtraction of the correlated background is performed the following way:
\begin{itemize}
  \item[-] The absolute correlated event rate per day is given by the reactor OFF data. The global correlation
with the muon rate is determined as accurately as possible, combining reactor ON and OFF data to constrain the
muon rate dependence. The final error band is shown as a grey shaded area in figure~\ref{fig:Muon_ONOFF}.
  \item[-] The muon rate is measured for each reactor ON period.
  \item[-] The corresponding correlated background and its associated uncertainty is deduced from the
correlation with the muon rate. This reactor OFF contribution is subtracted from the total rate of correlated events
measured reactor ON.
\end{itemize}

As the difference between average reactor ON and OFF muon rate is only \SI{0.5}{Hz} and the slope of the correction
is \SI{-22.8(50)}{event.day^{-1}.Hz^{-1}}, the correction reaches \SI{11.4}{event/day} and introduces a small systematic
uncertainty on the final neutrino rate of \SI{2.5}{event/day}. When the data sample is separated in shorter time period,
this error increases for each single data point, depending on the actual muon rate.


\subsection{Reactor induced correlated background}
\label{sec:reactor_bkg}
\subsubsection{Neutron induced correlated background}
Due to the proximity of the detector with the reactor, fast neutrons generated by the core could penetrate
the polyethylene shielding producing additional correlated background. This kind of events could be even more
problematic with respect to the cosmic ray induced neutrons since the associated background cannot be subtracted
using reactor OFF data. We consider here only one correlated background process: neutron elastic scattering
on nuclei mimicking the prompt event. Even if a fission generates several fast neutrons, it is extremely unlikely
that 2 neutrons from the same fission scatter off nuclei in the (small) Nucifer detector located several meters away.
We also consider that the inelastic scattering on nuclei process is negligible with respect to elastic neutron scattering
(see section~\ref{sec:cosmic_bkg_ori}).

\begin{figure}[h!]
\centering \includegraphics[width=1 \linewidth]{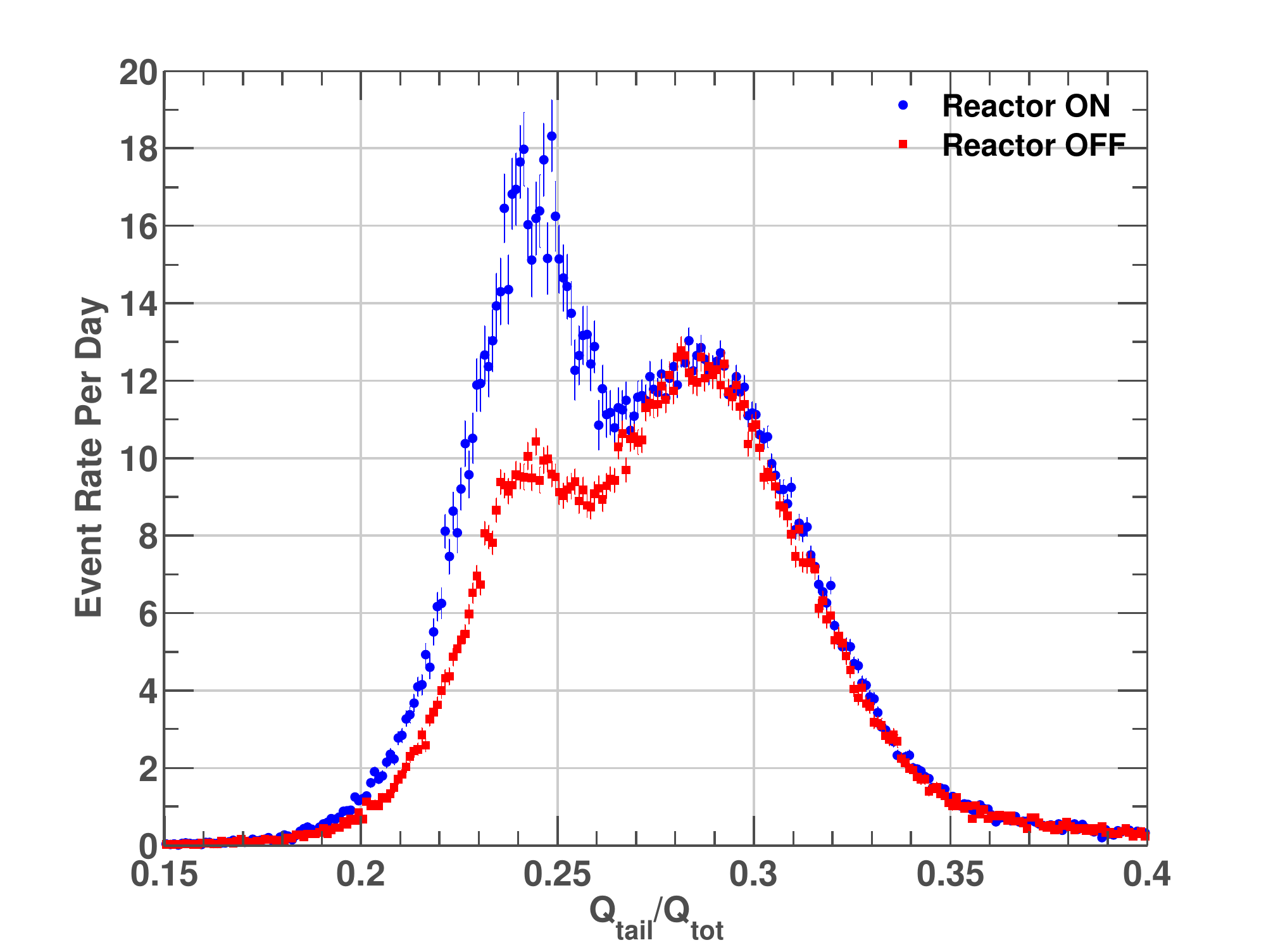}
\caption{\label{fig:PSDplot} Distributions of the PSD parameter \psd{} for reactor OFF (squares) and reactor
ON data (circles), obtained after subtraction of the accidental background and correction for the cosmogenic rate
modulation.}
\end{figure}

In order to check if a background contribution remains on top on the reactor OFF data, we exploit the Pulse Shape
Discrimination (PSD) capability of the Nucifer liquid scintillator. Aside the total charge recorded
for each trigger (\qtot{}), a late charge is also recorded (\qtail{}, see~\ref{sec:ExpLayout}).
The distribution of the ratio \psd{} of all prompt candidates is displayed in figure~\ref{fig:PSDplot} for reactor ON
and OFF data. As expected, two peaks emerge, one centred at $\psd{} \simeq 0.24$ corresponding to electron recoil and
another one centred at a higher value,~0.29, corresponding to nucleus recoils induced by fast neutrons. In the
comparison between the ON and OFF PSD spectra the neutrino signal clearly shows up as a Gaussian distribution
centred at~0.24, as expected from the positron (electron-like) recoil. For large \psd{} values corresponding to nucleus
recoils, both spectra overlap and demonstrate that the background subtraction is well under control.

Adjusting each curve with the sum of two Gaussians, the compatibility of the nucleus recoil peaks in our errors bars is
confirmed. However, the reactor ON peak exhibits 18~more events than the reactor OFF peak, in the \num{0.3}~to~\num{0.4}
\psd{} range. We conclude that the reactor induced correlated event rate is \SI{0(18)}{events/day}, and add this as an
additional systematic uncertainty to the final neutrino rate measurement. It is worth noting that the delayed gate used
to record \qtail{} has suffered from an electronic jitter during the major part of our data taking, not constant in time,
before we could identify and correct the defective gate generator. This electronic jitter is our best hypothesis
to explain the mismatch between the nuclei recoil peaks ON and OFF.

To better constrain the neutron induced correlated background, we performed a complete \tripoli{} simulation.
The starting point is a core simulation that samples outgoing fast neutrons ($>\SI{2}{MeV}$) at the surface of the core.
These particles are then used as a source for a second simulation that propagates the neutrons to the Nucifer target
volume, using sophisticated variance reduction techniques to push the particles through the thick absorbing media.
In particular, we used INIPOND, a special built-in module of \tripoli{} based on the exponential
transform method~\cite{CLARK}. This exponential biasing is performed using an importance map which provides information
on the probability, for each point of the phase space, for a particle to reach the detector. It is first calculated
with a simplified deterministic solver and then adjusted by hand. The code uses this information to adjust
the propagation of the neutrons along the path to the detector, thereby reducing the variance of the calculated
observables without introducing any bias.

We found that the neutron elastic scattering rate on hydrogen in the whole Nucifer target volume is \SI{4e-5}{event/day} 
for neutrons with energies higher than \SI{2}{MeV}, corresponding to an attenuation of the order of \num{e-27}.
This rate is a large overestimation because we neglected the real energy transfer, the quenching and the prompt-delayed
coincidence, considering each neutron elastic scattering on hydrogen as a background event satisfying all our selection cuts.
The statistical uncertainty reported by \tripoli{} is at the percent level but large systematic uncertainties have
to be considered, simply by the propagation of errors on initial parameters (cross-section, composition\ldots)
on such enormous attenuation. The results should therefore be considered as an order of magnitude estimate. Despite
the large uncertainty, our conservative approach ensures that the neutron induced correlated background is negligible.

Note that the separation between proton and electron-like recoils could in principle be used for further
rejection of the correlated background. A cut $\psd{}<0.27$ would remove about~2/3 of the reactor OFF
background. However this criterion is not currently in force in our analysis since the rather poor separation
between the 2~peaks implies a potentially large migration of events from both sides of the cut should
slight drifts in the pulse integration occur. The application of this PSD cut would thus lead to an extra
systematic uncertainty whereas our dominant background is coming from accidental coincidences.

\subsubsection{Gamma induced correlated background}
We also considered the gamma correlated background: a gamma ray with sufficiently high energy could excite a nucleus
by inelastic scattering, and if excited above the neutron separation energy the nucleus could emit a neutron.
This process demands very energetic gamma ray since the typical neutron separation energy is above \SI{10}{MeV}
for stable light nuclei (\nuc{C}{12}, \nuc{O}{16}, \nuc{Fe}{56}) and in the \SIrange{6}{8}{MeV} range
for heavy nuclei (Pb, Gd), with strong effects of neutron number parity (several MeV) so that some rare stable nuclei
have lower neutron energy separation (e.g. \SI{4.1}{MeV} for \nuc{O}{17}, \SI{4.9}{MeV} for \nuc{C}{13}).

If the gamma energy is high enough, the neutron could be fast and create correlated background as in the previous scenario.
But up to \SI{6}{MeV} proton recoils produce 3 times less light than electron recoil due to
quenching~\cite{O'Rielly1996745,vonKrosigk:2013sa}, therefore the neutron should have
at least \SI{6}{MeV} to enter our prompt energy window. Adding this energy to the neutron separation energy demands
an initial gamma energy higher than what is observed in the Nucifer spectrum (see Fig.~\ref{fig:spectra_single}).

Another possibility is the creation of a low energy neutron, the remaining energy being left to the gamma ray.
If both particles enter the detector and are absorbed, the gamma ray can create the prompt event and the neutron capture
the delayed event, with a time correlation identical to neutrino events. To enter the prompt energy windows, the gamma
ray must have kept at least \SI{2}{MeV}, and because the neutron would have low energy, the probability to go through
the polyethylene shielding without being captured on Boron would be negligible. Therefore the initial inelastic
scattering should take place in the tank, on some odd neutron number light nuclei.
But such nuclei have low natural concentration (percent or less), at such energy the cross-section of ($\upgamma,n$)
reaction (\SI{2.0e-4}{barn} for \nuc{C}{13} at \SI{7}{MeV}~\cite{Soppera2014294,Koch:1976qlm}) is much lower
than Compton scattering cross-section (\SI{3.96e-1}{barn} for C at \SI{7}{MeV}~\cite{NIST-XCOM})
and the interaction rate of high energy gamma in Nucifer is low enough (about \SI{0.5}{Hz} above \SI{7}{MeV})
so that this process can be safely neglected.

\subsection{Detection efficiency and associated systematics}
\label{sec:DetEffAndSyst}
\subsubsection{Detection efficiency}
\label{sec:DetEff}

As explained in section~\ref{sec:ExpLayout}, a dedicated software package generates a set of simulated \neb{}
events in the same format than the data. To determine the detection efficiency we apply the exact same
analysis chain and compute the amount of rejected events. Since the path lengths of $\upgamma$-rays and
neutrons in the liquid scintillator ($\sim$ \SI{20}{cm}) are not negligible with respect to the size of the
detector target (\SI{1.2}{m} diameter), edge effects induce sizeable correlations between the analysis
selection criteria. In consequence the total detection efficiency is not the simple product of the
efficiencies of all single cuts. 
It is defined as the ratio of the number of simulated events passing all the analysis
selection criteria to the total number of neutrino vertices generated in the liquid scintillator. Using the selection
procedure described previously the global detection efficiency is found to be \EffPC{}.

As shown on table~\ref{tab:det_eff}, taking each cut individually the selection of the delayed energy deposition
(\SI{47.9}{\%} efficiency) is the selection criteria that degrades most the efficiency. This is due to the lack of containment of
the multiple high-energy $\upgamma$-rays released after the neutron capture on Gd in the small, \SI{619.2}{mm} in radius
and \SI{704.9}{mm} in height, detector target. The time selection is the next limiting cut (\SI{63.1}{\%}), rejecting
the same amount of event below and after the gate, and finally the low energy cut at \SI{2}{MeV} has a rather strong
influence on the prompt event selection (\SI{76.6}{\%}). According to the simulation, IBD neutrons are captured mostly
by the Gadolinium (\SI{85}{\%}), then by the Hydrogen (\SI{11}{\%}), the Boron (\SI{2}{\%}, in the polyethylene shielding)
and the remaining by the steel components. \SI{7.2}{\%} of IBD neutrons are captured outside the target.

The efficiency of the multiplicity cut is computed by counting the number of diode triggers with and without
applying this cut. Because of their different single rates reactor ON and OFF have different multiplicity efficiencies
found to be~0.974 and~0.990, respectively. They both have a negligible associated systematic uncertainty
and the correction is applied during the data analysis. Finally the \SI{3.9}{\%} dead-time induced by the $\upmu$-veto
is independently computed with negligible uncertainty from the $\upmu$~triggers, taking in account overlaps between
veto gates. This veto time is subtracted from the total live-time of the analysed runs.

\begin{table}[!ht]
\begin{center}
    \begin{tabular}{l>{\centering} m{2.5cm}  @{}m{0pt}@{}}
    \toprule
      Selection cut 				& Efficiency	&\tabularnewline \colrule
      $\SI{6}{\us}<\Delta t<\SI{40}{\us}$	& \SI{63.1}{\%}	&\tabularnewline[2pt]
      $\SI{2.0}{MeV}<\Ep{}<\SI{7.1}{MeV}$	& \SI{76.6}{\%}	&\tabularnewline
      $\SI{4.2}{MeV}<\Ed{}<\SI{9.6}{MeV}$	& \SI{47.9}{\%} &\tabularnewline \colrule
      $\epsilon_{\text{det}}$ 			& \EffPC{}	&\tabularnewline[2pt] \colrule  \colrule
						& Dead-time	&\tabularnewline[2pt] \colrule
      Multiplicity ON				& \SI{2.6}{\%}	&\tabularnewline[2pt]
      Multiplicity OFF				& \SI{1.0}{\%}	&\tabularnewline
      Muon veto					& \SI{3.9}{\%}	&\tabularnewline
     \botrule
\end{tabular}
\end{center}
\caption{\label{tab:det_eff} Detection efficiency or induced dead-time of the selection cuts. As explained in the text
the total efficiency is not the simple product of cut efficiencies because of correlations between the cuts.}
\end{table}

\subsubsection{Systematic uncertainties}
\label{sec:Systematics}
\begin{figure}[h!]
\centering
\includegraphics[width=1 \linewidth]{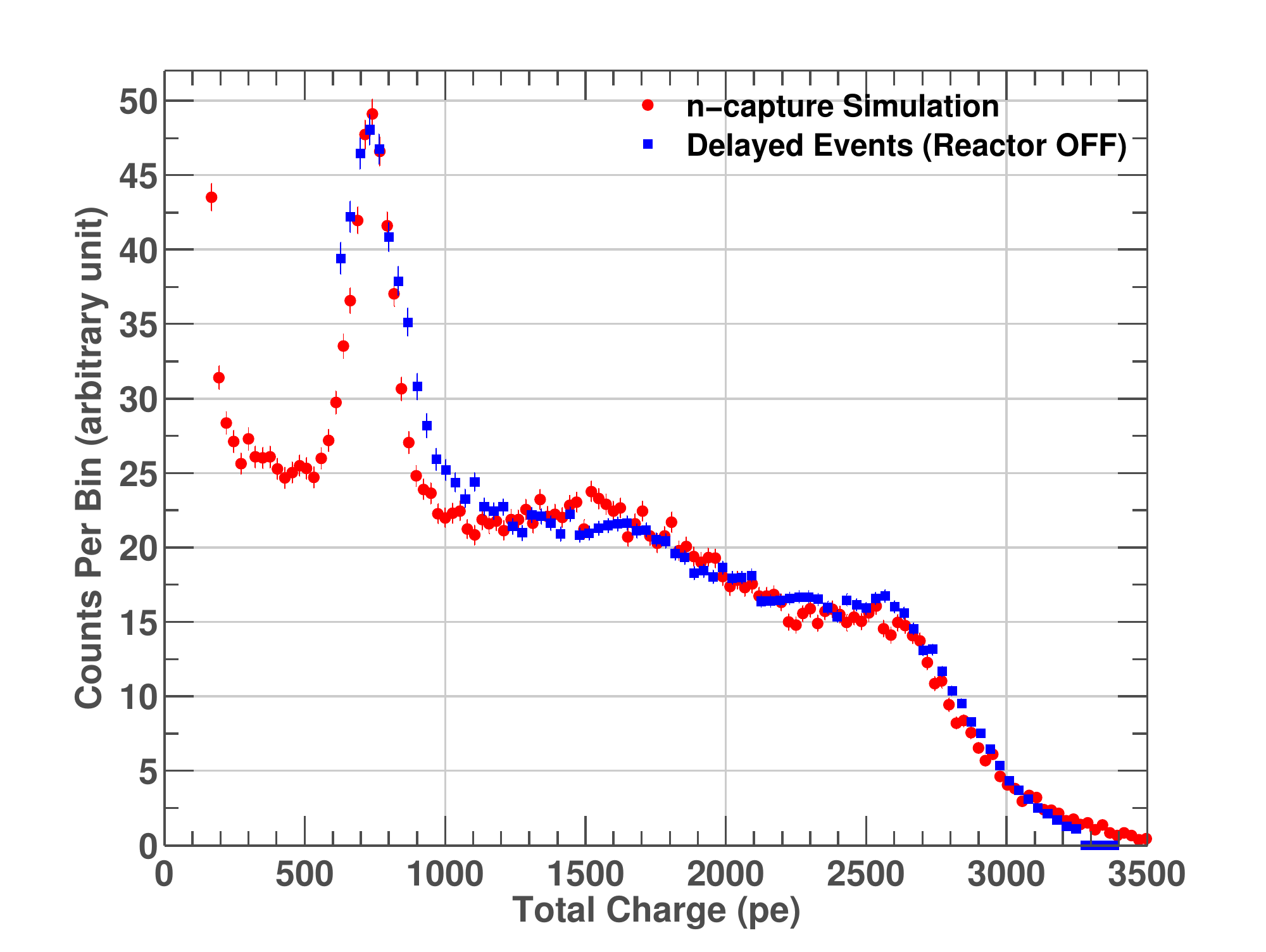}
\caption{Photo-electron spectrum of delayed energy events during a reactor OFF run compared to a Geant4.9.4
simulation of low energy neutrons uniformly distributed in a sphere of \SI{2}{m} radius centred on the target
vessel. The n-capture peak at \SI{750}{PE} (\SI{2.2}{MeV}) is prominent because of the numerous captures
occurring in the polyethylene shielding around the target. The data threshold comes from the software analysis,
thus not affecting the $n(\mathrm{H},\mathrm{D})\upgamma$ peak, while the simulation has no threshold.}
\label{fig:energy_neutron}
\end{figure}

The uncertainty on the energy scale is determined by comparing the delayed spectrum of correlated pairs in
the reactor OFF data (Fig.~\ref{fig:energy_neutron}) with the Geant4.9.4 simulation of thermal neutron uniformly
distributed inside the lead shielding, once the Geant4 simulation has been tuned to reproduce at best the source spectra
at different positions. This set of data was chosen because it is independent from the calibration data, it shows
several spectral features among the whole energy range and the energy depositions are uniformly distributed in the whole
target volume, as expected from neutrino candidates. Note that the source in the simulation, uniformly distributed
thermal neutron, is a crude approximation of the natural neutron flux at shallow depth. But the simulation is
not intended to reproduce exactly the measured spectrum, particularly the relative weight of the energy features,
but to allow the comparison of the position of these features to establish the energy scale uncertainties.

The main features of the energy spectrum (n-capture peak on H at \SI{750}{PE}, a middle bump at about \SI{1700}{PE}
and a high energy edge of n-capture on Gd) are clearly visible for both data and simulation. The $n(\mathrm{H},
\mathrm{D})\upgamma$ peak position, the transition between the two plateaux at \SI{1200}{PE} and \SI{2200}{PE},
and the $n(\mathrm{Gd},\mathrm{Gd})\upgamma$ shoulder position are fitted on both data and MC. The largest difference is
found to be \SI{2.5}{\%} and is taken as a safe estimate of the energy scale uncertainty.
To determine the efficiency uncertainty due to energy scale, we repeated our analysis of the simulated neutrino data set
1000~times with a variation of energy cuts. At the beginning of each analysis, a factor is shot in a Gaussian
distribution of relative standard deviation \SI{2.5}{\%}, and each energy cut is multiplied by this factor.
The distribution of all the neutrino rates found is a Gaussian which standard relative deviation, \SI{1.0}{\%},
is the uncertainty on efficiency due to energy scale.

Nucifer is made of a unique volume of liquid scintillator enclosed in a stainless steel tank. Therefore,
neutrino interaction outside the target leading to light production can only happen in the acrylic buffer,
with the positron depositing at least \SI{2}{MeV} in the liquid scintillator. Thus only the first
centimetres of the buffer are concerned. We simulated IBD events uniformly distributed in the buffer and found that
less than \SI{0.15}{\%} of such events pass the analysis cuts, a negligible quantity (less than \SI{0.5}{event/day}).

The uncertainty associated to the time selection criteria is determined by comparing Geant4.9.4, Geant4.9.4 with
the addition of a module dedicated to thermal neutron physics~\cite{Collin:2014} and \tripoli{} simulations,
using different description of low energy neutron physics. We find a \SI{1.7}{\%} error from the dispersion
of these 3 simulation results, taking as central value the Geant4.9.4 simulation with the additional thermal
neutron module. We also add a \SI{1}{\%} uncertainty on the efficiency due to the modelling of the gamma cascade
following a neutron capture on Gd, by comparing the results of our 3 simulations.

The measurement of the neutron mean capture time with the AmBe source or the reactor OFF correlated events can not be
directly used because the initial energy and spatial distributions of the neutrons are different.
Indeed, the thermalization phase of neutrons from IBD is not detectable due to our electronic dead time
(\SI{6}{\us}), while the $\Delta t_{\upgamma-n}$ distribution of AmBe events shows a peak at \SI{10}{\us},
then an exponential decay with a shorter parameter of \SI{17.6(12)}{\us} for a source at the centre of the detector.
Neutrons from a\nuc{Cf}{252} source would suffer from the same problem.

Finally, the efficiency and the associated uncertainty are \EffwithErrPC{} (see table~\ref{tab:eff_sys}),
but here \AbsErrEffPC{} is an absolute uncertainty on the efficiency of \EffPC{}. Therefore, the relative uncertainty
on efficiency is $\AbsErrEffPC/\EffPC=\RelErrEffPC$, and the predicted neutrino detection rate is
$\Rnu{pred} = \SI{913}{\neb{}/day} \times \EffwithErrPC = \RatePred{}$ (see section for the predicted \neb{}
interaction rate~\ref{sec:parameters}).

\begin{table}[!ht]
\begin{center}
    \begin{tabular}{l>{\centering} c @{}m{0pt}@{}}
    \toprule
      Source				& Uncertainty		&\tabularnewline \colrule
      Time cut				& \SI{1.7}{\%}		&\tabularnewline[2pt]
      Energy scale			& \SI{1.0}{\%}		&\tabularnewline
      Gd(n,$\upgamma$) cascade		& \SI{1.0}{\%}		&\tabularnewline
      IBD in buffer			& \SI{0.2}{\%}		&\tabularnewline
    \colrule
      Absolute efficiency uncertainty	& \AbsErrEffPC{}	&\tabularnewline[2pt]
    \colrule \colrule
      Relative efficiency uncertainty	& \RelErrEffPC{}	&\tabularnewline[2pt]
      \neb{} interaction rate		& \SI{4.6}{\%}		&\tabularnewline
    \colrule
      Total prediction uncertainty	& \SI{8.5}{\%}		&\tabularnewline[2pt]
    \botrule
\end{tabular}
\end{center}
\caption{\label{tab:eff_sys} Breakdown of the detection efficiency systematic uncertainties,
and final prediction uncertainty. \neb{} interaction rate uncertainties are listed on table~\ref{tab:flux_sys}.}
\end{table}

\section{Results}
\label{sec:results}

\begin{figure}[h!]
\centering \includegraphics[width=1 \linewidth]{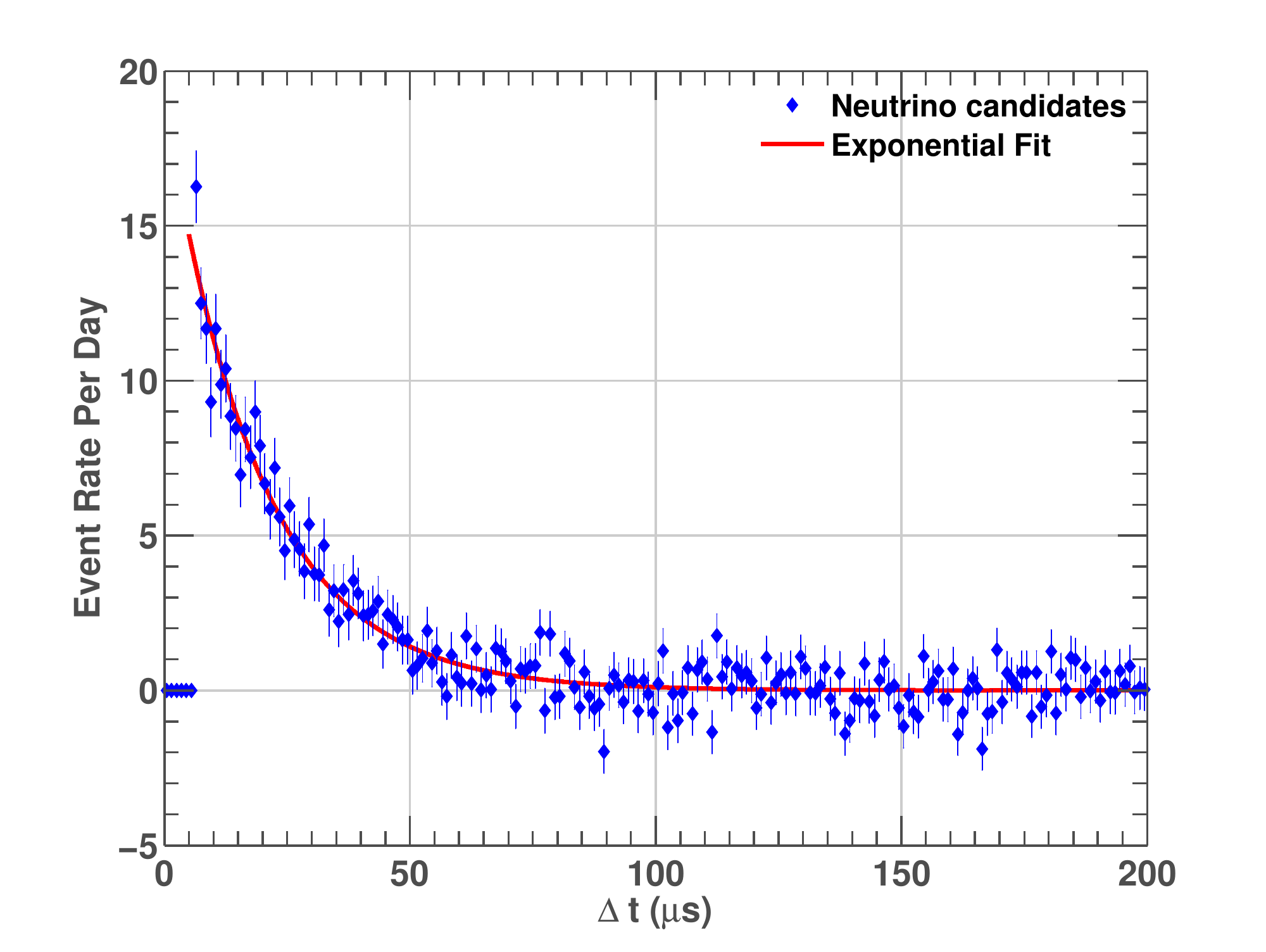}
\caption{\label{fig:Deltat_Neutrinos} Distribution of the prompt-delayed time differences for all neutrino candidates
after background subtraction. The spectrum of the prompt-delayed time differences is found to be compatible with
a pure exponential curve with a decay time of \SI{19.7(9)}{\us}, in agreement with the expected value.}
\end{figure}

The distribution of neutrino events is obtained after subtracting the accidental background and the correlated
background from the raw number of candidate pairs. In Fig.~\ref{fig:Deltat_Neutrinos} the spectrum of the
prompt-delayed time differences is found to be in agreement with a pure exponential curve with a decay time of
\SI{19.7(9)}{\us}, compatible with the decay time expected from the simulation.

We can use this curve to set a constraint on possible residual accidental background that would
essentially show up as an extra offset. Adding this free parameter in the fit the value converges on
\num{8e-3} with an uncertainty of \num{6.6e-2} that makes it perfectly compatible with zero.
Integrating this error bar between~6 and \SI{40}{\us} we obtain a possible ``accidental-like'' contribution
of \SI{2.3}{events/day}. This corresponds to only \SI{0.07}{\%} of the measured accidental rate and thus
demonstrates the good quality of the subtraction.

From the number of entries we get a total of \SI{40760}{\neb} detected in Nucifer, corresponding to a mean
rate of $\Rnu{obs} = \RateObs{}$. The expected detected neutrino rate is $\Rnu{pred} = \RatePred$
and therefore $\Rnu{obs}/\Rnu{pred} = \RateObsToPredRatio$.

\begin{table}[h!]
\begin{center}
  \begin{tabular}{l@{~~}c@{~~~}c}
  \toprule
    Single rates (Hz)           & ON			& OFF			\tabularnewline
  \colrule
    $\upmu$ veto                & $344.3$		& $376.4$		\tabularnewline
    $\upmu$ veto \& Saturation  & $110.6$		& $110.1$		\tabularnewline
    Singles                     & $177.1$		& $65.7$		\tabularnewline
    Prompt singles              & $75.4$		& $16.1$		\tabularnewline
    Delayed singles             & $15.7$		& $1.6$			\tabularnewline
  \colrule \colrule
    Pairs (/day)		& ON			& OFF			\tabularnewline
  \colrule
    Candidates			& \RateCandONnum{}	& \RateCandOFFnum{}	\tabularnewline
    Accidentals			& \RateAccONnum{}	& \RateAccOFFnum{}	\tabularnewline
    Correlated			& \RateCorrONnum{}	& \RateCorrOFFnum{}	\tabularnewline
  \colrule
    \Rnu{obs}			& \multicolumn{2}{c}{\RateObs{}}		\tabularnewline
    \Rnu{pred}			& \multicolumn{2}{c}{\RatePred{}}		\tabularnewline
  \botrule
  \end{tabular}
\end{center}
 \caption{\label{tab:rates} Summary of relevant rates for reactor ON and OFF periods. Note that the number of
observed neutrinos \Rnu{obs} is not directly the ON-OFF difference of correlated events, as correction
for the multiplicity cut and for the different mean muon rates have been applied.
Errors reported on candidate and accidental event rates are statistical only, while errors on correlated rates also
show systematic uncertainties due to background subtraction. All other systematic uncertainties are applied
on the predicted rate.}
\end{table}

\begin{figure*}[t!]
\centering \includegraphics[width=0.65 \linewidth]{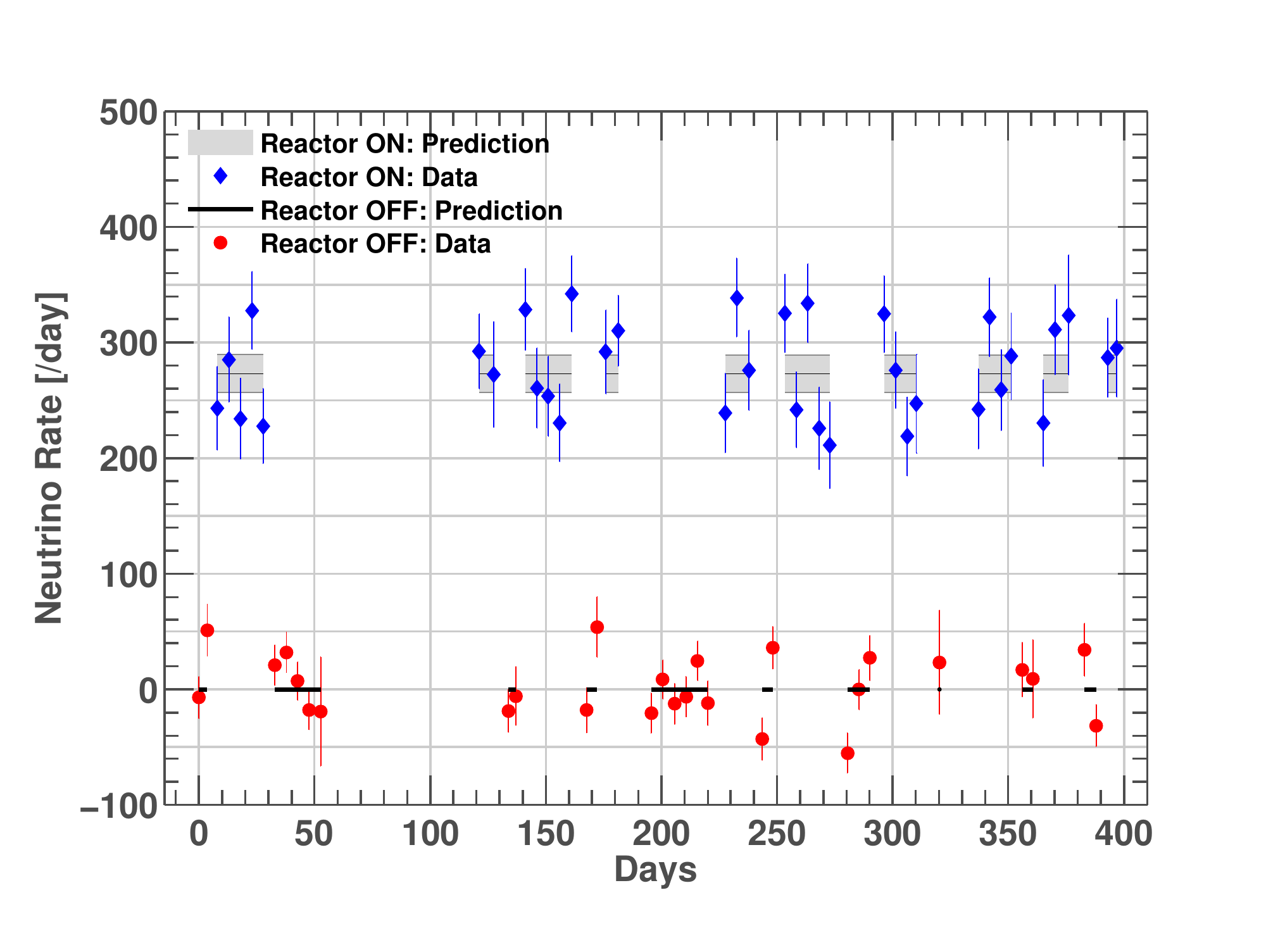}
\caption{\label{fig:Evolution} Antineutrino rate measurement monitoring the Osiris nuclear reactor operations.
Each data point (blue diamonds) corresponds to about \SI{5}{days} of data taking with its associated
statistical uncertainty. The grey shaded area is the rate expectation above the mean correlated background
when the reactor is OFF, referred as zero level here and plotted as red dots.}
\end{figure*}

All relevant rate, averaged over the full reactor ON and OFF data sets, are summarized in table~\ref{tab:rates}.
The time evolution of the detected neutrino rate is shown in figure~\ref{fig:Evolution} where the data are
grouped in periods of about \SI{5}{days}. The alternation of ON-OFF periods is clearly visible. The null rate
corresponds to the correlated background event rate averaged over the full reactor OFF data.

The statistical accuracy per \SI{5}{days} of data taking is about \SI{11}{\%}. As expected the intrinsic statistical
accuracy of the daily neutrino rate is decreased by a factor~4 due to the total background contribution being
16~times higher than the signal (\AccToSignalRatio{}~from the accidentals and \CorrToSignalRatio{}~from the
cosmic-ray-induced correlated events). As illustrated in figure~\ref{fig:FissionFractions} the isotopic composition of
the Osiris reactor is quite stable in nominal operation. Consequently only sub-percent variation of the emitted
\neb{} flux are expected between the beginning and the end of a cycle~\cite{VanMinh:2013}. The observation
of this Osiris intra-cycle rate evolution is clearly out of reach of a Nucifer-like detector.

\begin{figure}[h!]
\centering \includegraphics[width=1 \linewidth]{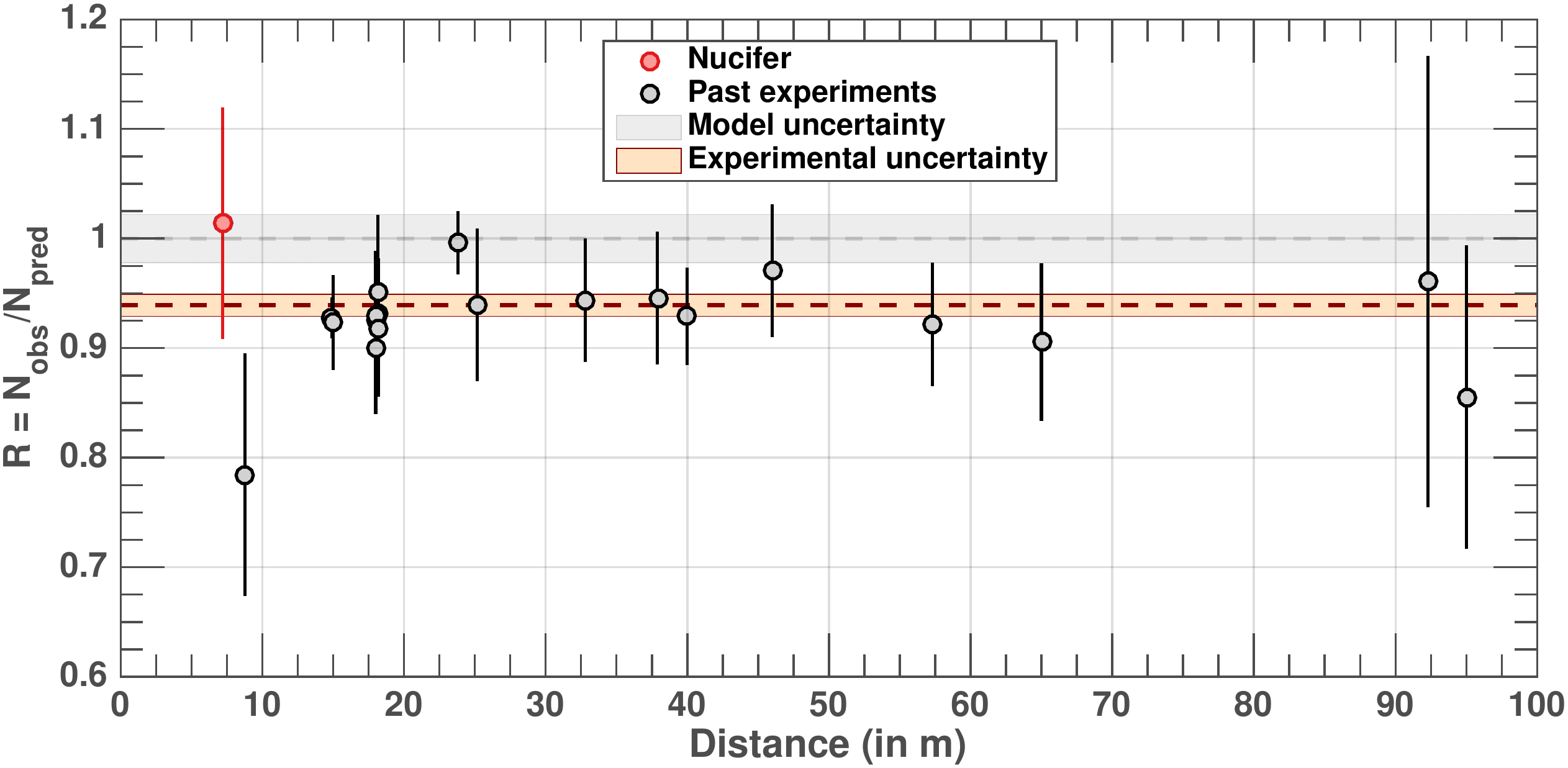}
\caption{\label{fig:RAA_Nucifer} Comparison of the Nucifer measurement of $\Rnu{obs}/\Rnu{pred}$ with others
short baseline experiments and the average value of the RAA~\cite{Mention:2011rk}. Without Nucifer, the average ratio
is \num{0.938(24)} (\SI{2.6}{\upsigma}), and with Nucifer the average is \num{0.940(24)} (\SI{2.5}{\upsigma}).}
\end{figure}

The current statistical sample accumulated by the Nucifer experiment provides a modest sensitivity to test
the Reactor Antineutrino Anomaly. The current \RelErrEffPC{} large uncertainty on the absolute predicted
normalization factor prevents any strong conclusion in favour or against the averaged $\Rnu{obs}/\Rnu{pred}$
discrepancy, as shown by figure~\ref{fig:RAA_Nucifer}.
Improving these results is beyond the scope of this publication.
A significant work will be necessary to refine the prediction of the expected neutrino rate, including an improved
determination of efficiency, and to reduce the reactor induced background systematics.

\section{Sensitivity to the plutonium content of the core}
\label{sec:Pu_mass}
With the end of the Cold War, hundreds of tons of weapon-grade plutonium were determined to be surplus to
U.S. and Russian defence needs. In April~2010 the US and Russian governments signed a protocol amending the
2000~Plutonium Management and Disposition Agreement (PMDA), which commits each country to dispose of no
less than 34~metric~tons (MT) of excess weapon-grade plutonium and envisions disposition of more weapon-grade plutonium
over time. The combined amount, \SI{68}{MT}, represents enough material for several thousands of nuclear weapons.
The current approach is to transform the weapon-grade plutonium into mixed oxide fuel and
irradiate it in reactors. We study here how a small neutrino detector like Nucifer could monitor this
kind of operation to guarantee that plutonium is really being burnt in the reactor.

The neutrino rate measured by Nucifer is used as a calibration point for a nuclear fuel highly enriched in
\Urfive{} with \SI{92}{\%} of the fissions coming from this isotope (see section~\ref{sec:Signal}).
This approach can be seen as a measurement of our absolute normalization with the statistical \SI{2.5}{\%} accuracy
(from \RateObs{}) provided by our sample.

\begin{figure}[h!]
\centering \includegraphics[width=1 \linewidth]{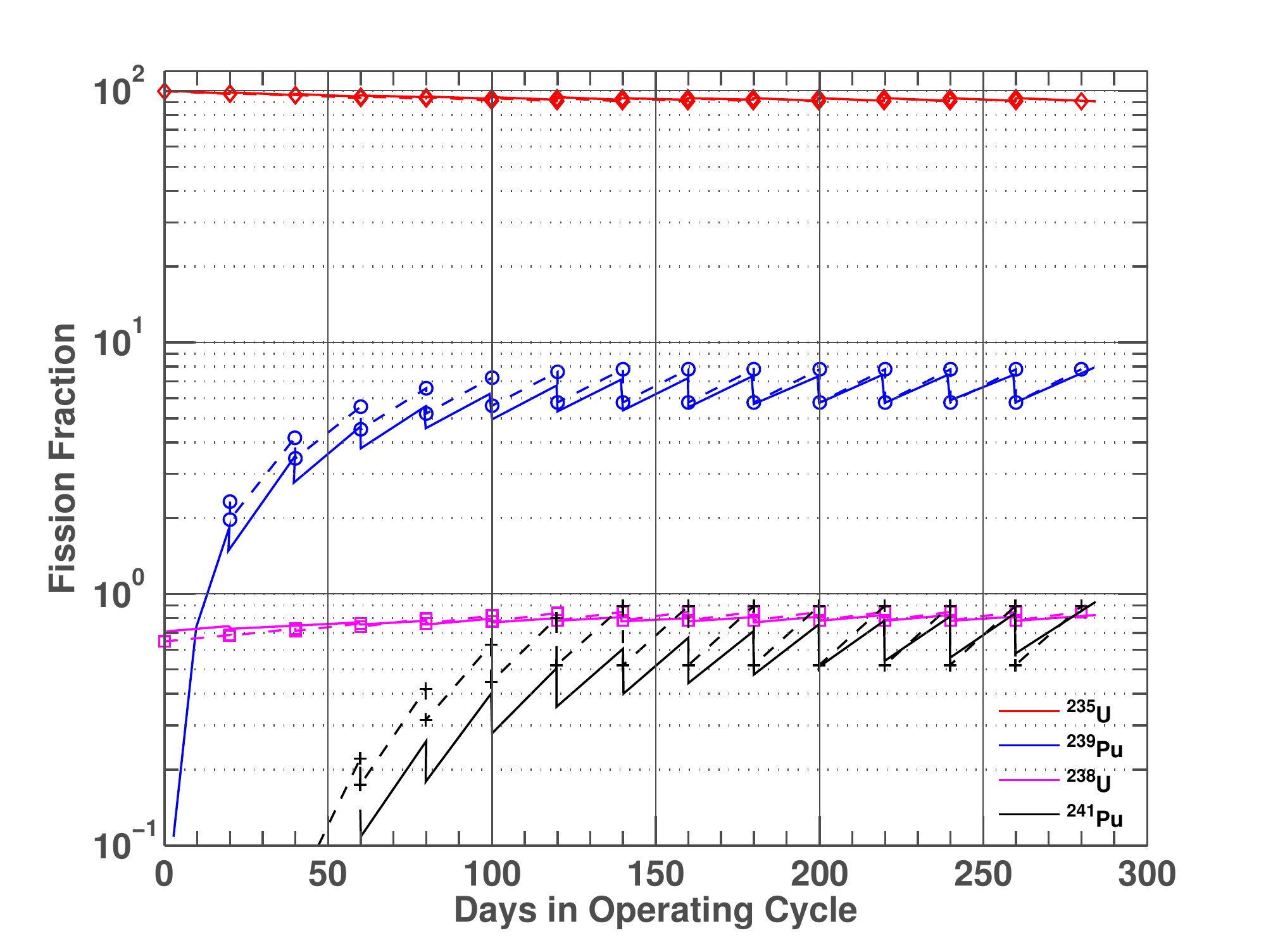}
\caption{Fission fraction in~\% of each fissile isotopes, \Urfive{}, \Ureight{}, \Punine{}, and \Puone{}, as
predicted by the MURE simulation (solid lines)~\cite{VanMinh:2013}. Here the core is initially filled with fresh fuel
and its isotopic composition evolves for 14 cycles, but all actual Osiris cycles correspond to the equilibrium regime.
The dotted lines illustrate the effective approach dedicated to study the sensitivity to the \Punine{} mass
in the core after it is tuned to reproduce the reference MURE results.}
\label{fig:FissionFractions}
\end{figure}

Using the MCNP Utility for Reactor Evolution (MURE)~\cite{MURE:Proc, MURE:NEA}, we were able to simulate
the evolution of the Osiris fuel with a full 3D Monte-Carlo simulation.
Figure~\ref{fig:FissionFractions} shows an example of the evolution of the fission fractions of \Urfive{},
\Ureight{}, \Punine{}, and \Puone{} as predicted by MURE~\cite{VanMinh:2013}. Starting from a fresh uranium
fuel enriched at \SI{19.75}{\%} an equilibrium is reached after the $7^{th}$ cycle. As expected the fission of the
\Urfive{} dominates by far and all fission fraction evolutions during a cycle are moderate.

We then simulated the operation of the Osiris reactor with part of its fuel elements loaded with \Punine{},
like in UOX-MOX fuel. The MURE simulation of the fuel evolution being very time consuming,
we used for this purpose a simple model of the core~\cite{Joubert:2009}: we solved the Bateman's equations
of the evolution of the uranium and plutonium isotopes assuming a uniform neutron flux adjusted to match the
mean Osiris thermal power and with effective nuclear cross sections (n-capture and fission) calibrated to reproduce
the MURE simulation results. Figure~\ref{fig:FissionFractions} shows that a good agreement could be reached for the
nominal core composition of Osiris, good enough for the purpose of this study.

\begin{figure}[h!]
\centering \includegraphics[width=1 \linewidth]{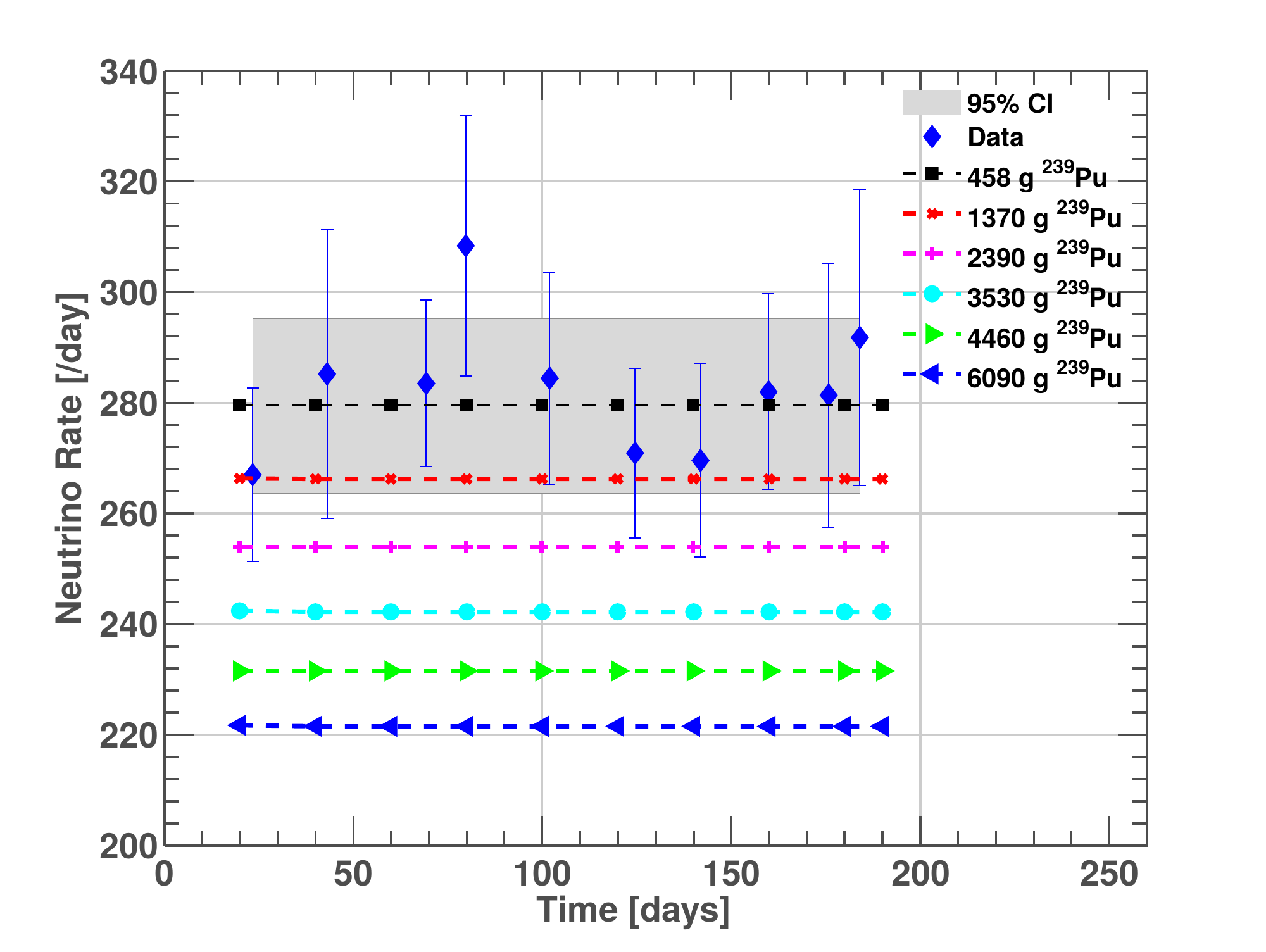}
\caption{\label{fig:PuSensitivity} Mean neutrino rate detected by Nucifer (blue diamonds), concatenating Osiris cycles.
The grey area shows the interval at \SI{95}{\%}~C.L. The other horizontal lines illustrate the predicted evolution of the
detected flux for an increasing mass of Plutonium in the Osiris core.}
\end{figure}

In standard operation the Osiris core is loaded with \SI{14.00(75)}{kg} of \Urfive{} and \SI{450(50)}{g} of
\Punine{}, the error bar representing the variation of the masses around these mean values during a
\SI{21}{days} cycle. This equilibrium is reached after an initial load of about \SI{20}{kg} of \Urfive{}. To study
the impact of plutonium content on the detected neutrino flux we tried various initial configurations with
up to \SI{10}{kg} of Pu in the initial composition of the core. However the initial mass of all fissile materials
(mainly \Urfive{} and \Punine{}) was always kept equal to \SI{20}{kg} for all configurations. Then we simulated the
reactor core evolution during several cycles at constant full power, until the equilibrium was reached. At this
stage we determined the associated mean mass of \Punine{} in the core from the number of atoms predicted
by our model. Finally, the mean neutrino rate was computed from the number of fissions per isotope in our
model and from the reference ratios of detected \neb{} per \Urfive{} fission listed in table~\ref{tab:NuRatePerIsotope}.

\begin{table}[h!]
\begin{center}
  \begin{tabular}{l>{\centering}m{1.cm}>{\centering}m{1.cm}>{\centering}m{1.cm}>{\centering}m{1.cm}@{}m{0pt}@{}}
  \toprule
	Isotope			& \Urfive{}	& \Ureight{}	& \Punine{}	& \Puone{}	& \tabularnewline[2pt]
  \colrule
	Relative \neb{} rate	& 1		& 1.512		& 0.635		& 0.900		& \tabularnewline[2pt]
  \botrule
  \end{tabular}
\end{center}
 \caption{\label{tab:NuRatePerIsotope} Variation of detected antineutrino rate per fission normalized to one fission of
\Urfive{}. The reference antineutrino spectra per fission are taken from~\cite{Huber:2011wv} and~\cite{Haag:2013raa}
and interaction cross-section from~\cite{Strumia:2003zx}. The prompt energy cuts of the Nucifer analysis are applied
corresponding to $\SI{2.78}{MeV} < \anuen < \SI{7.88}{MeV}$.}
\end{table}

As expected the neutrino rate decreases as the mass of \Punine{} in the core, hence the contribution
of \Punine{} fissions, increases. This is illustrated in figure~\ref{fig:PuSensitivity} where a global normalization
factor is applied to our model in order to set the prediction for the nominal core composition equal to the observed
rate. Then assuming the same data taking period and background conditions, the detection of a lower \neb{} rate with
\SI{95}{\%} confidence level is equivalent to a drop of 2.23~times of our current rate uncertainty~(7.1), i.e.~15.8~less
detected \neb{}~per day. This sensitivity limit is reached for about \SI{1.5}{kg} of \Punine{} in the core,
a mass representing \SI{10}{\%} of the total mass of fissioning elements in the Osiris core at equilibrium. As already
stated before, this sensitivity is driven by the large accidental background at the Osiris site. If
the same Nucifer detector was installed close to a commercial reactor with a baseline (\SI{25}{m}) and background
conditions (S/B=4) similar to the SONGS experiment~\cite{Songs:2007}, the same sensitivity
could be reached in less than \SI{3}{days}.

\section{Conclusion and outlook}
This article reported on the features and performances of the Nucifer experiment operating at the Saclay research
centre of the French Alternative Energies and Atomic Energy Commission (CEA) since Spring~2012.
The experimental configuration, the Osiris research nuclear reactor, and the detector setup have been presented.

We discussed the installation and the operation of the detector \Dist{} away from the Osiris reactor, making
Nucifer the second world-shortest baseline neutrino experiment ever being operated. Being at such shot distance
of the core lead to great difficulty in mitigating and controlling reactor induced accidental background.
The high-accuracy measurement and subtraction of the huge gamma-ray reactor induced background was validated, as well
as a novel method for assessing cosmic ray induced backgrounds at very shallow depth, correcting for the evolution
of the atmospheric pressure during reactor OFF/ON periods.

Eventually reactor antineutrinos at very short baselines can be used to probe the Reactor Antineutrino Anomaly
and to search for possible oscillations into sterile neutrino species.
However, because of the current lack of precision in the emitted neutrino flux as well as the large accidental
background level, no definitive conclusion could be obtained yet. A more refined evaluation of the expected neutrino rate
would be necessary through further improvements of the detector response modelling, which are beyond the scope of this
paper. Nevertheless Nucifer data can be now included into global analyses and it could perhaps constrain some
of the still-allowed oscillation scenarios at short baselines. Dedicated experiments are clearly necessary to confirm
or reject the short baseline oscillation hypothesis, such as the Stereo~\cite{Helaine:2016bmc},
Neutrino-4~\cite{Serebrov:2015ros}, PROSPECT~\cite{Ashenfelter:2015uxt}, or SOX~\cite{Borexino:2013xxa,Machulin:2015eca}
experiments expected in the next months or years.

The detection efficiency was improved by a factor~of~3 with respect to past reactor neutrino experiments
dedicated to nuclear safeguards, reaching a mean daily rates of about \SI{300}{\neb{}/day}
with a cubic-meter scale target volume. Therefore, the achieved statistics is large enough to be of interest for
cooperative monitoring regimes. Nucifer has also shown a great stability, suitable for relative monitoring
of the neutrino rate within a few percent, and safe automatic operation for a few years with only little occasional
maintenance. Hence, this operation regime approaches the requirements of any IAEA monitoring apparatus, making
a Nucifer-like antineutrino detector suitable to monitor and safeguard nuclear reactors.


As a first societal application, and within the framework of the Plutonium Management and Disposition Agreement for
the disposal of Weapon-grade plutonium, we illustrated the possibility of monitoring the plutonium content
in the Osiris core by showing that Nucifer could detect the presence of about \SI{1.5}{kg} of Pu at \SI{95}{\%}~C.L,
based on the actual data. This result would be improved with Nucifer deployed further away from a more powerful core,
by intrinsically lowering the dominant gamma-ray accidental background component.

\section*{Acknowledgements}
We are indebted to the management and staff of CEA/DEN/Osiris reactor for their full support towards the realization
of the Nucifer experiment and the excellent support they provided to the Nucifer collaboration during the detector
operation; we acknowledge the support the CEA/DSM, CEA/DAM, CEA/DEN, CNRS/IN2P3, and the Max Planck Gesellschaft. We
would like to thank Gabriele Fioni for initiating the collaboration between CEA physicists and the OSIRIS reactor staff.



\bibliography{nucifer}

\end{document}